\documentclass[10pt,conference]{IEEEtran}

\usepackage{framed}
\usepackage{cite} 
\usepackage[table]{xcolor}
\usepackage{graphicx}
\usepackage{booktabs}
\usepackage{caption}
\usepackage{subcaption}
\usepackage{amssymb}

\usepackage{hyperref}
\hypersetup{
    colorlinks,
    linkcolor={red!50!black},
    citecolor={blue!50!black},
    urlcolor={blue!80!black}
}

\usepackage{pifont}
\newcommand{\cmark}{\ding{51}}%
\begin{document}

\title{Relating Reading, Visualization, and Coding for New Programmers: A Neuroimaging Study}

\makeatletter
\newcommand{\linebreakand}{%
  \end{@IEEEauthorhalign}
  \hfill\mbox{}\par
  \mbox{}\hfill\begin{@IEEEauthorhalign}
}
\makeatother

\author{
\IEEEauthorblockN{Madeline Endres}
\IEEEauthorblockA{Computer Science and Engineering\\
University of Michigan\\
endremad@umich.edu}
\and
\IEEEauthorblockN{Zachary Karas}
\IEEEauthorblockA{Department of Psychology\\
University of Michigan\\
zackar@umich.edu}
\and
\IEEEauthorblockN{Xiaosu Hu}
\IEEEauthorblockA{Department of Psychology\\
University of Michigan\\
xiaosuhu@umich.edu}
\linebreakand
\IEEEauthorblockN{Ioulia Kovelman}
\IEEEauthorblockA{Department of Psychology\\
University of Michigan\\
kovelman@umich.edu}
\and
\IEEEauthorblockN{Westley Weimer}
\IEEEauthorblockA{Computer Science and Engineering\\
University of Michigan\\
weimerw@umich.edu}
}


\maketitle


\begin{abstract}

Understanding how novices reason about coding at a neurological level has implications for training the next generation of software engineers. In recent years, medical imaging
has been increasingly employed to investigate 
patterns of neural activity associated with coding activity. 
However, such studies have focused on advanced undergraduates
and professionals. In a human study of 31 participants,
we use functional near-infrared spectroscopy to 
measure the neural activity associated with introductory
programming. In a controlled, contrast-based experiment,
we relate brain activity when coding 
to that of reading 
natural language or mentally rotating objects 
(a spatial visualization task). 
Our primary result is that all three tasks---coding, 
prose reading, and mental rotation---are mentally distinct
for novices. 
However, while those tasks are neurally distinct, 
we find more significant differences between prose and
coding than between mental rotation and coding. 
Intriguingly, we generally find
more activation in areas of the brain associated with spatial ability 
and task difficulty for novice coding compared to that reported in studies with
more expert developers.
Finally, in an exploratory analysis, we also find a neural activation
pattern predictive of programming performance 11 weeks later. 
While preliminary, these findings both expand on
previous results (e.g., relating expertise to a similarity
between coding and prose reading) and also provide a
new understanding of the cognitive processes underlying
novice programming.



\end{abstract}

\pagestyle{plain} 


\section{Introduction}
\label{sec:introduction}

In recent years, the software engineering community has increasingly used medical neuroimaging to understand the cognitive processes behind programming~\cite{siegmund2014understanding, FSE17, floyd2017decoding, nakagawa2014quantifying, fakhoury2018effect, huang19, krueger2020}. Unlike eye-tracking or other biometric methods, neuroimaging  pinpoints brain regions activated while completing specified tasks. Many of the neuroimaging studies in software engineering have compared programming to reading or spatial manipulation, two skills with well-understood cognitive structures. As broadly summarized in Table~\ref{tab:introTable}, such studies have generally found striking similarities between code comprehension and prose reading~\cite{siegmund2014understanding, FSE17, floyd2017decoding}, and one study has found similarities between spatial reasoning and data structure manipulation~\cite{huang19}.

Neuroimaging studies of programmers have the potential to improve our understanding of expertise, to inform software engineering pedagogy, and to guide tool development and retraining (see Floyd \emph{et al.}~\cite[Sec.~II-D]{floyd2017decoding} for a summary). Critically, however, to the best of our knowledge all of the software engineering neuroimaging studies thus far have only studied programming \emph{experts} that are either professionals or students with multiple years of experience
(e.g.,~\cite[Sec.~3.3]{FSE17}). Tantalizingly, one study~\cite{floyd2017decoding} found that coding became more neurologically similar to reading for programmers with even greater expertise. However, to realize the potential of neuroimaging for understanding software engineering expertise, we must also directly observe true \emph{novices}. Studying novice programmers is critical for exploring how cognitive processes for coding evolve. In this paper, we seek to close this gap by presenting the first neuroimaging study of novice programmers during code comprehension (cf.~\cite{floyd2017decoding, FSE17}).

Studying novice programmers presents several challenges relative to studying experts. First, we must create experimental stimuli that are amenable for novices with little to no previous coding experience; the coding stimuli in previous neuroimaging studies all involve constructs (e.g., trees~\cite{huang19}) or software engineering tasks (e.g., code review~\cite{floyd2017decoding}) unfamiliar to novices. Second, we must pay particular care to recruit participants with equivalent programming expertise; even in an introductory course, some students have substantially more programming experience than others. Finally, we propose and use an experimental protocol that follows up with participants months later to assess their programming progression using a written assessment; to our knowledge, no previous neuroimaging studies in software engineering have involved 
time-delayed outcome measurements.

\begin{table}[t]
\belowrulesep 0px   
    \centering 
    \begin{tabular}{l|ccc} 
            Experiment          & Like reading?  & Like spatial & Novices?\\
             &                                    & reasoning? &\\
        \toprule
        Siegmund \emph{et al.} (2014)~\cite{siegmund2014understanding} &\cellcolor{green!25}\cmark&&\cellcolor{yellow!25}\textbf{?} \\
        Siegmund \emph{et al.} (2017)~\cite{FSE17} &\cellcolor{green!25}\cmark &&\cellcolor{yellow!25}\textbf{?}\\
        Floyd \emph{et al.} (2017)~\cite{floyd2017decoding} & \cellcolor{green!25}\cmark&&\cellcolor{yellow!25}\textbf{?}\\
        Huang \emph{et al.} (2019)~\cite{huang19} & &\cellcolor{green!25}\cmark&\cellcolor{yellow!25}\textbf{?}\\
    \end{tabular}
    \caption{``What is coding like in programmers' brains?'' We use this informal summary of selected previous work to motivate and contextualize the experiments in this paper.}
    \label{tab:introTable}
\end{table}

To better understand the cognitive processes of novice programmers, we use functional near-infrared spectroscopy (fNIRS) to conduct a controlled neuroimaging study with 31 participants with no previous programming experience, all enrolled in the same introductory computing class. We conduct these scans during the first third of the class. We compare participants' brain activation patterns while coding to those while reading prose or using spatial reasoning (i.e., mentally rotating 3D objects). We also use a written assessment at the end of the semester to conduct a preliminary exploration of an aspect of learning, assessing if novices' brain activation patterns while coding can predict their future programming ability.

We find that, for novices, coding is a working memory intensive task that is neurally distinct from both a spatial task and prose reading ($p < 0.01$, $q < 0.05$). Unlike previous work with experts which generally reports strong similarities between coding and reading~\cite{floyd2017decoding, siegmund2014understanding,FSE17}, we observe more significant and substantial differences between coding and reading than we do for coding and spatial tasks. This indicates that novices rely heavily on visiospatial cognitive processes while coding. Finally, we observe that particular activation patterns at the beginning of a course can predict how well students perform on a final programming assessment 11 weeks later; in general, the \emph{less} similar the activation patterns are between coding and mental rotation, the \emph{better} an individual performs ($r = 0.48$, $p = 0.006$). This may indicate that novices who use more problem-solving intensive strategies at the beginning of the semester (i.e., find programming more challenging) make less progress. We also compare our results to those of previous neuroimaging studies with more expert software engineers, and we close with a discussion of the implications of our results on introductory programming pedagogy and future software engineering research.

\section{Background}
\label{sec:background}

We give an overview of material necessary for understanding the methods and experiments in this paper. In Section~\ref{sec:medicalImaging}, we discuss neuroimaging and fNIRS, and in Section~\ref{sec:notation}, we present relevant notation. In Section~\ref{sec:spatialReasoningCognition}, we discuss the cognition behind spatial ability, and in Section~\ref{sec:readingCognition}, we discuss the cognition behind reading. See Section~\ref{sec:related} for a discussion of work more directly related to software engineering.

\subsection{Neuroimaging and fNIRS}
\label{sec:medicalImaging}

Brain activity and cognitive processes
can be studied using 
\emph{functional neuroimaging} techniques.
We focus on functional near infrared spectroscopy (fNIRS). It is non-invasive, avoids the ionizing radiation present in other methods (e.g., PET, CT), and can measure
activity in brain regions not accessible to 
some invasive techniques (e.g., electrocorticography). 
Importantly, fNIRS offers higher
spatial resolution than EEG, and higher temporal resolution than fMRI, which is important for studies
relating a brain region's contribution to a specific task. Finally, fNIRS can be
used in more natural and ecologically-valid
environments (e.g., standard desktop computer use, etc.) compared to alternatives
like fMRI (which requires participants to lie still in a small tube and also complicates the use of
keyboards~\cite{krueger2020}). These properties motivate
our decision to use fNIRS for a study of
novice software engineers. 

fNIRS makes use of the 
\textit{hemodynamic response}, or change
in neuronal blood flow to active brain regions, to measure brain activity~\cite{buxton2004modeling}. fNIRS
measures this via the use
of near-infrared light: transmitters
and receivers are placed on a ``cap'' worn by
participants. Oxygen-rich and oxygen-poor blood
have different light absorption properties, 
so the hemodynamic responses in a given brain region between a transmitter
and receiver pair (referred to as a \emph{channel}) can be measured over time. 
fNIRS  measures  concentration  changes  in  such oxygenated  and deoxygenated  blood.
The number of fNIRS publications had doubled every 3.5 years since 1992~\cite{boas2014twenty}, 
and it has been used to study human 
development, injury,
and psychiatric conditions~\cite{boas2014twenty,lloyd2010illuminating,ehlis2014application,obrig2014nirs}.

For our purposes, the use of fNIRS imposes
two key experimental constraints:
contrast-based design and task duration. 
First, fNIRS experiments typically involve participants completing tasks (e.g., mentally rotating
objects or solving programming problems)
while time-series data is recorded. Carefully
controlled experiments
are necessary, in which the activity 
observed during
one task is contrasted against the activity
observed during another. 
This allows confounding brain activations
(e.g., motor cortex activity from moving the
lungs to breathe) to be eliminated from consideration. 

Second, because fNIRS is based on 
the hemodynamic response, care must
be taken to model~\cite{scicurious_2012,bennett2009neural} the onset of
neuronal blood flow (which peaks slightly after
stimuli are presented~\cite{henson2002detecting,aamand2013no}) and the design must avoid
saturation and weaker signals for
tasks involving long activity~\cite{lindquist2009modeling}.
As a result, the hemodynamic response can 
typically be studied only in experiments
with brief stimuli (e.g., under 30 seconds
per question). Furthermore, fNIRS is only able to penetrate a few centimeters down into the brain. More specifically, the near-infrared light can reach a depth that is roughly half the distance between a transmitter-receiver pair, depending on the wavelengths and light intensities used~\cite{ferrari2012brief}. 


Despite these limitations, fNIRS specifically,
and medical imaging in general, are
growing in popularity for use in 
software engineering studies
(e.g.,~\cite{siegmund2014understanding, floyd2017decoding,nakagawa2014quantifying,ikutani2014brain,fakhoury2018effect,Duraes16,Castelhano2018,FSE17,Peitek:2018:ESEM,huang19,krueger2020}).
They provide a physically-grounded insight
into cognition without relying on potentially unreliable self-reporting~\cite{huang19}.

\subsection{Neuroscience Vocabulary and Notation}
\label{sec:notation}

\noindent\textbf{Vocabulary:} The cerebrum of the human brain is composed of two (largely symmetric) hemispheres, left and right, and four primary lobes: frontal, temporal, parietal, and occipital. Loosely, the \emph{frontal lobe} is at the front of each hemisphere, the \emph{temporal lobe} is on the side of each hemisphere, the \emph{parietal lobe} is at the top of each hemisphere, and the \emph{occipital lobe} is at the back of each hemisphere. Activation is called \emph{bilateral} if both hemispheres are activated and \emph{lateralized} if one hemisphere activates disproportionately. Throughout this paper, we will use various schema to refer to locations on the cerebrum's \emph{cortex}, the brain's outer layer of neural tissue. One such schema is \emph{Brodmann Areas}, an anatomical classification system for the cortex~\cite{brodmann2007brodmann}. Broadmann areas (BAs) divide the cortex into 52 bilateral regions based on architectural neurological features. Many BAs also have associated neurological functions. For instance, BAs 41 and 42 are associated with auditory processing. We will also sometimes refer to regions by their common names or location on a lobe. Three important regions mentioned in this paper are Wernicke's area (left hemisphere, back of the parietal lobe), Broca's area (left hemisphere, lower frontal lobe), and the dorsolateral prefrontal cortex  or DLPFC (bilateral in the frontal lobe). Wernicke's and Broca's areas are strongly associated with language functions while the DLPFC is associated with working memory.

\vspace{2mm}

\noindent\textbf{Notation:} In this paper, we will use the neuroimaging notation \emph{Task A $>$ Task B} to indicate the contrast between brain activation patterns for two experimental tasks. The results of these contrasts are reported as statistical $t$-values corresponding to each fNIRS channel. These $t$-values range from $-8$ to $+8$; a positive $t$-value indicates that the specified brain area was \emph{more} active during Task A than Task B, while a negative $t$-value indicates that the specified brain area was \emph{less} active during Task A than Task B. Values closer to $-8$ and $+8$ represent stronger activation contrasts between the two tasks, and only areas with significant contrast ($p < 0.01$) after correction for multiple comparisons ($q < 0.05$) are reported. Finally, we note that contrast tests are directional: a significant contrast in \emph{Task A $>$ Task B} does not imply that the inverse contrast \emph{Task B $>$ Task A} will also produce significant results. This is because the differences in the inverse contrast may be too small to be statistically significant.

\subsection{Spatial Reasoning and Cognition}
\label{sec:spatialReasoningCognition}

We now turn to a discussion of Spatial Reasoning and its neurological representations. \emph{Spatial Reasoning} refers to an individual's general ability to mentally manipulate objects and encompasses skills such as mental rotation, mental folding, pattern recognition, and spatial perception~\cite{MargulieuxSpatialTheory}. Spatial reasoning has been shown to correlate with performance in a variety of activities including mathematics\cite{hegarty1999types, wai2009spatial}, general engineering~\cite{sorby2018does}, and programming~\cite{jones2008spatial, parkinson2018investigating}. Spatial ability is also malleable and can be improved through training~\cite{uttal2013malleability}. 

Part of spatial reasoning, mental rotation involves the imagined rotation of a two- or three-dimensional object around an axis in three dimensional space~\cite{zacks2008neuroimaging}. Figure~\ref{fig:mentalStim} depicts one of the mental rotation stimuli we used (see Section~\ref{sec:fNIRSStimuli}). The difficulty of a mental rotation problem is determined by the size of the angle of rotation; Shepard and Metzler found that the time a participant took to solve a given problem increased linearly with respect to the angle of rotation between the two objects~\cite{shepard1971mental}. In this paper, we use mental rotation ability as a validated proxy for more general spatial reasoning ability.

Generally, neuroimaging has found that mental rotation activates the posterior parietal and occipital cortices (BAs, 7, 17, 19, 39, and 40---see Zacks~\cite{zacks2008neuroimaging} for a survey). While bilateral, the parietal and occipital activation tends to be slightly stronger in the right hemisphere; the right parietal lobe in particular is believed to be important for spatial ability and spatial attention tasks~\cite{culham2001neuroimaging}. Many mental rotation neuroimaging studies have also revealed bilateral activation in the supplementary motor cortex, an area associated with motor control and planning (BA 6)~\cite{zacks2008neuroimaging}. This frontal activation is most common for mental rotation tasks that allow motor stimulation strategies. 

\subsection{Reading and Cognition}
\label{sec:readingCognition}

Next, we include a brief summary of the neurological processes associated with reading. Neuroimaging has revealed that language is supported by a complex network of cognitive areas generally lateralized to the left hemisphere (see Price~\cite{price2012review} and Vigneau \emph{et al.}~\cite{vigneau2006meta} for surveys). Some language processes are localized to specific structures in the brain, while
other language processes arise from a distributed network of areas with multiple functions~\cite{price2012review}. 

Two key left-hemisphere brain areas associated with reading are Broca's area and Wernicke's area. Located in the frontal lobe, Broca's area (BAs 44 and 45) plays an important role in language production and, to a lesser extent, language comprehension~\cite{vigneau2006meta}. Wernicke's area is in the posterior temporal lobe and is associated primarily with comprehension of both spoken and written language~\cite{vigneau2006meta}. 

\section{Experimental Setup and Design}
\label{sec:setup}

We now present our experimental design for understanding the neurological basis for novice programmers.\footnote{A replication package containing all of our recruitment and stimuli materials can be found at \url{https://github.com/CelloCorgi/ICSE_fNIRS2021}.} Our experiment was conducted in two parts: an initial fNIRS scan and a written followup assessment. During the fNIRS scan, participants were shown three types of stimuli: code comprehension, mental rotation, and prose reading. During the written post-test, participants completed a validated language-agnostic programming assessment. All participants were enrolled in the same 15-week introductory programming course. The initial fNIRS scans were held during the first third of the semester while the written post-test was held during the last week of the semester. This design allows the controlled exploration of the relationships and contrasts between reading, coding, and spatial reasoning for novice programmers. It also opens a preliminary investigation of neurological factors that might be predictive for programming. In the rest of this section we describe our recruitment protocol (subsection~\ref{sec:recruitment}); outline our fNIRS data collection, experimental setup, and stimuli (subsections~\ref{sec:fNIRSData_Collection} and~\ref{sec:fNIRSStimuli}); and describe our followup programming assessment (subsection~\ref{sec:posttest}).

\subsection{Participant Recruitment}
\label{sec:recruitment}
Participants were all recruited from the same CS1 course at the University of Michigan, a large public US university, via a combination of email, forum posts, and an in-class presentation. To be eligible, participants had to be over 18, have no prior programming experience, only be enrolled in that one programming course, and have be available to attend the study. 

Recruitment occurred during the third week of the fifteen-week semester, and the fNIRS scans were carried out in the third through fifth week of the semester. All participants were thus in the first third of their first programming course while completing the scan. Participants received $\$20$ in compensation. In total, we collected fNIRS data from 37 participants. Data from 31 participants passed our analysis quality threshold (see Section~\ref{sec:analysis}). The final 31 participants (24 female, 7 male) ranged in age from 18 to 21. 

Beyond the initial fNIRS scan, we also followed up with participants via email at the end of the semester, inviting them to attend an additional written programming assessment. Of our 31 participants, 23 participated in this written post-test. Post-test participants were compensated an additional $\$20$. 

During participant selection, we were keenly aware that confirming the absence of previous programming ability for a diverse population is a challenging task. To mitigate self-selection bias, we implemented checks for participant programming ability in three places: recruitment, pre-screening, and statistical validation of scores on a written programming test.
During recruitment, both in-person and written, we emphasized that participants could have no prior programming experience of any kind. Prospective participants were explicitly told, in person, that even minimal practice or exposure to textual or visual programming languages (such as Scratch~\cite{scratch} or MIT App Inventor) counted as prior experience. 

During prescreening, participants indicated if they ``had any prior programming experience'' with one of ``Yes'', ``No'', or ``Other/unsure''. We only retained participants who selected ``No'' outright. The presence of the ``Other/unsure'' option mitigates some self-selection bias in experience reporting. We also asked potential participants to indicate concurrent and previous course enrollment from a list of courses at our university. Courses that contained ``programming'' or ``code'' in their syllabus or description precluded study participation. These questions eliminated 18\% of pre-screening respondents.

Finally, at the same time as the demographics questionnaire, participants were given a brief programming test. The test consisted of twelve pseudocode multiple-choice questions and is a validated measure of CS1 concepts~\cite{ParkerSCS1}. We expected low scores: novices should have no prior programming exposure (beyond the first few weeks of their current course). Indeed, we found an average score of 23\% (random guessing yields 20\%). There was no statistically significant difference between participant scores and a random distribution. We believe that these three mitigations help account for self-selection bias issues and give confidence that our pool contains novices (but acknowledge that the issue is reduced, rather than eliminated). 

We also observe that unusually for software engineering studies, a majority of our population (77\%) were female. We believe that the high ratio of females in our population is caused by a combination of: a more gender-balanced population pool than most software engineering studies, fNIRs data signal, and self selection bias. First, the course we recruited from is fairly gender balanced: around 45--50\% of students identify as women, compared to 22\% for CS overall at our university. Second, by chance, two-thirds of the participants who we were unable to analyze due to fNIRS signal quality (i.e., measurement noise) were male. Third, women often volunteer for college studies at higher rates than men and for different reasons~\cite{lobato2014impact} while men are more likely to have some prior programming experience~\cite{sackrowitz1996unlevel} and thus be excluded from our study.

\subsection{fNIRS Data Collection and Setup}
\label{sec:fNIRSData_Collection}

Each participant's fNIRS data was collected during a single session which lasted 1.5 hours. First, participants gave informed consent and filled out a short demographic survey. Next, participants watched a training video preparing them for the scan, a 30--45 minute process that involved fitting each participant for the fNIRS cap by moving hair to optimize sensor contact with the scalp, and thus, signal quality. 

During the fNIRS scan, the participant sat in a chair facing a monitor wearing the fNIRS cap. The room was kept dim to reduce the amount of ambient light that could interfere with the fNIRS data collection. Participants were also instructed to stay as still as possible. Each participant was shown 90 stimuli: 30 mental rotation stimuli, 30 reading-based stimuli, and 30 coding-based stimuli. All stimuli asked the participant to choose one of two answers. Participants indicated their answer by pressing a corresponding key on a standard keyboard. For each stimuli, participants had up to 30 seconds to respond. The 90 stimuli were in randomized order and were broken into three blocks of 30 questions, each containing 10 stimuli of each type. Between stimuli, participants were shown a fixation cross for 2--10 seconds. Between blocks, participants had an optional longer break to rest and/or drink water. 

We now present technical information about our fNIRS device and cap. We used a CW6 fNIRS system (TECHEN, Milford, MA, USA)
with 690 nm and 830 nm wavelengths. It has fiber-optic cables which transmit light from the device to sensors connected to the participant's cap. These sensors are either transmitters that emit light or receivers that detect light. As a result, fNIRS collects a participant's hemodynamic response only along a set of pre-defined channels between transmitters and receivers. The location, number, and coverage of these channels is determined by the cap design. 
We used a probe configuration similar to that used in Huang \emph{et al.}~\cite{huang2019distilling}
with dense coverage of transmitter-receiver pairs in the occipital, frontotemporal, and
frontal regions. In total, 15 receiver and 30 transmitter fibers were used, yielding 55
channels from which data were collected, covering Broadmann areas 6--9, 17--19, 21, 22, 39--41, and 44--47. 
The data from these channels were then analyzed using both NIRS Brain AnalyzIR Toolbox~\cite{nirs-brain-analyzir} and custom scripts written in MATLAB.
A picture of our cap coverage is in Figure~\ref{fig:capDiagram}; it includes areas identified in previous mental rotation studies as well as key language areas. 

Our number of channels is higher compared to many previously published fNIRS papers in software engineering, giving broader coverage~\cite{ikutani2014brain, nakagawa2014quantifying}. We used two cap sizes to accommodate different head circumferences (58 cm and 60 cm). Signals were sampled at 50 Hz.

\begin{figure}[t] 
    \centering 
    \includegraphics[width=0.8\linewidth]{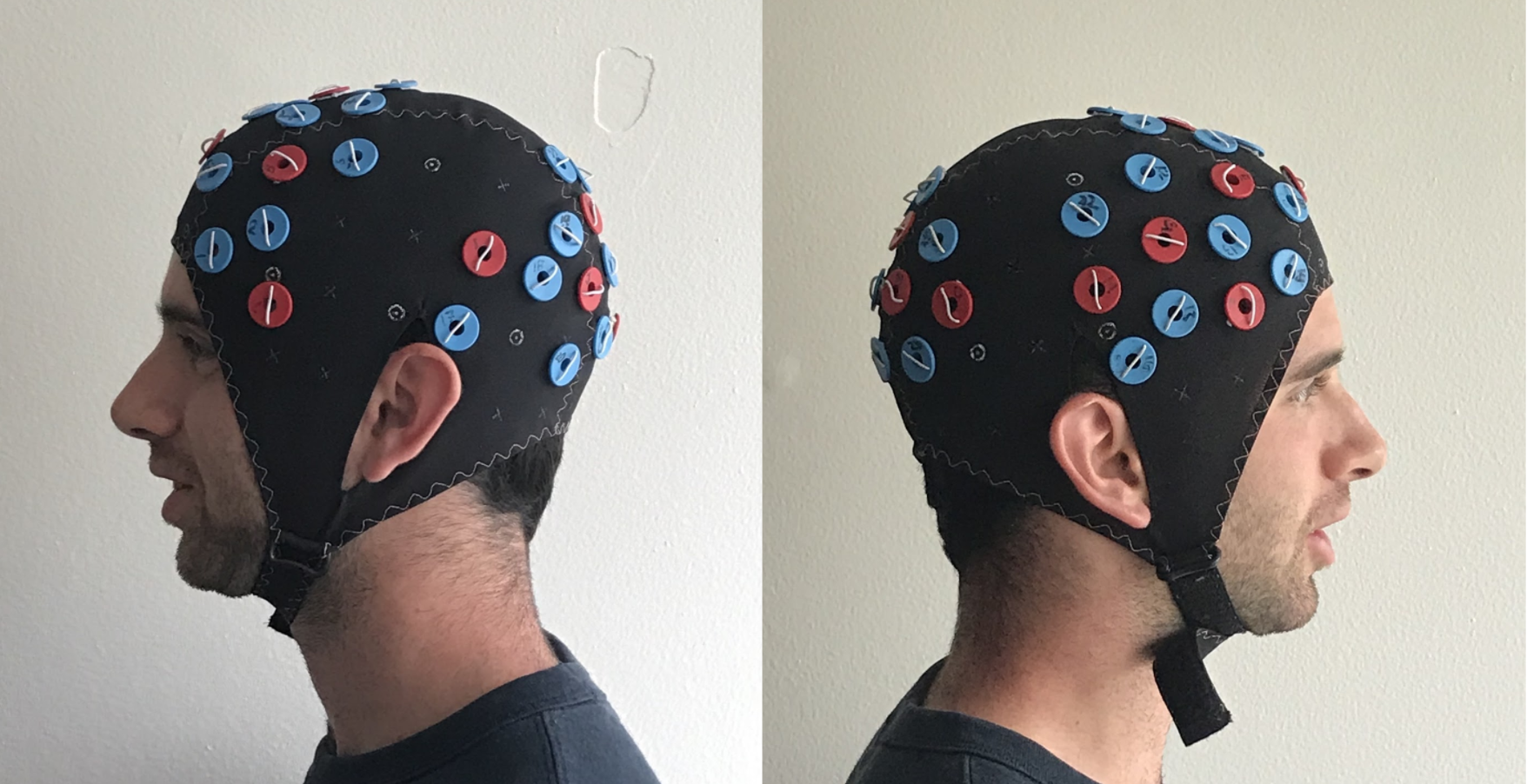}
    \caption{fNIRS cap used in our experiments. Red circles are light transmitters while blue circles are light receivers. fNIRS is only able to observe brain activation on channels between nearby transmitters and receivers.}

    \label{fig:capDiagram}
\end{figure}

\subsection{fNIRS Stimuli}
\label{sec:fNIRSStimuli}
We now turn to a description of the content of our fNIRS stimuli. As mentioned in Section~\ref{sec:fNIRSData_Collection}, participants were shown three categories of stimuli: mental rotation, reading-based stimuli, and coding-based stimuli. All three types asked the participant to choose between two answers \emph{A} and \emph{B}. 

We use mental rotation stimuli adapted from Peters and Battista's Mental Rotation Stimulus Library~\cite{peters2008applications}. These stimuli are designed to induce brain activation associated with mentally rotating 3D shapes, one facet of visiospatial cognition (See section~\ref{sec:spatialReasoningCognition}). In each mental rotation stimulus, the participant was asked to choose which of two objects was a possible rotation of a third object (see Figure~\ref{fig:mentalStim} for an example). To admit direct comparison with previous work, our mental rotation stimuli are the same stimuli used by Huang \emph{et al.} when analyzing data structure manipulation brain activation with experienced software developers~\cite{huang19}. 
\begin{figure*}
\begin{subfigure}[b]{0.3\textwidth}
    \centering
    \includegraphics[width=\linewidth]{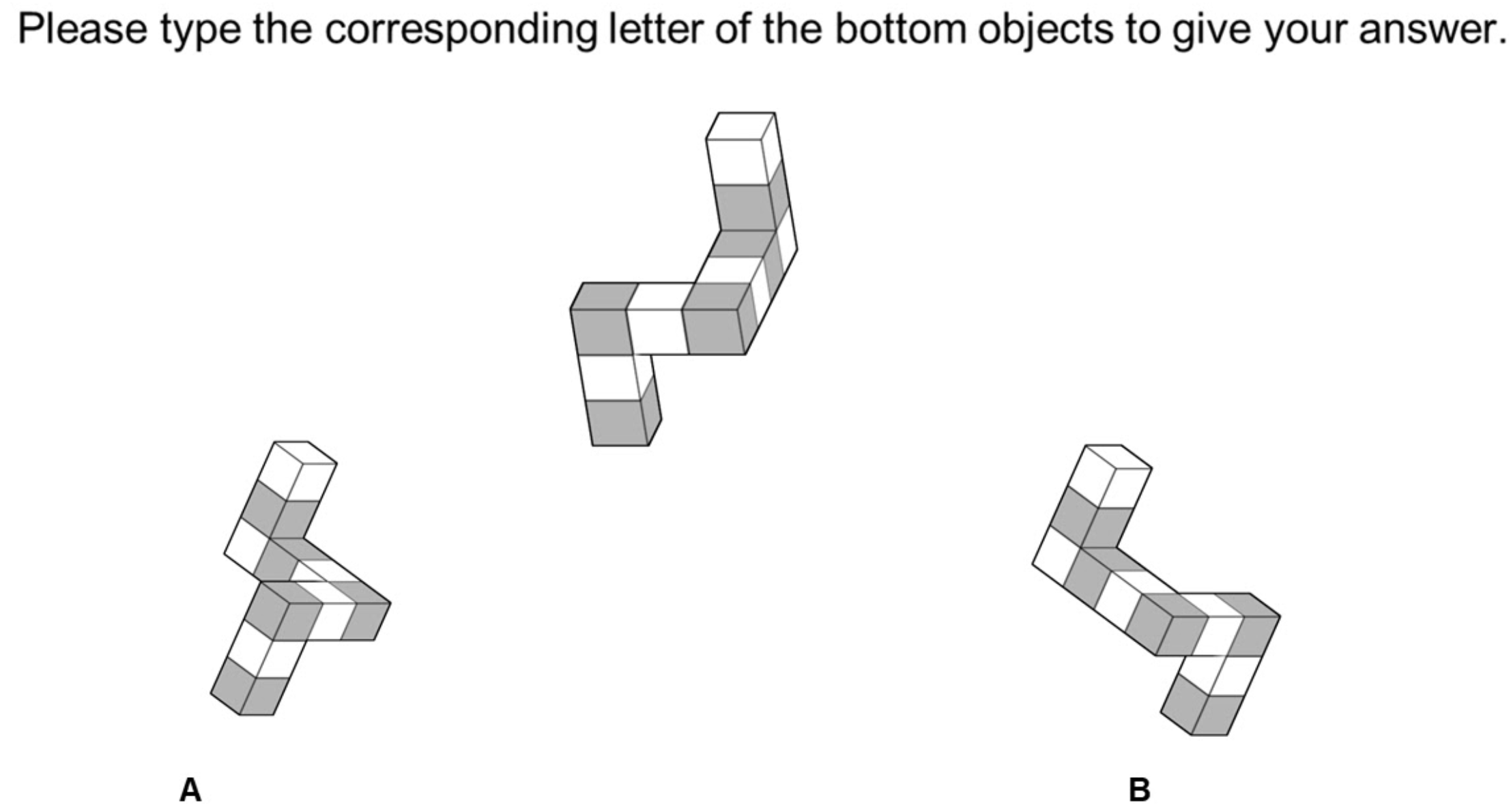}
    \caption{Mental Rotation: correct answer is ``A''.}
    \label{fig:mentalStim}
\end{subfigure}
\hfill
\begin{subfigure}[b]{0.3\textwidth}
    \centering
    \includegraphics[width=\linewidth]{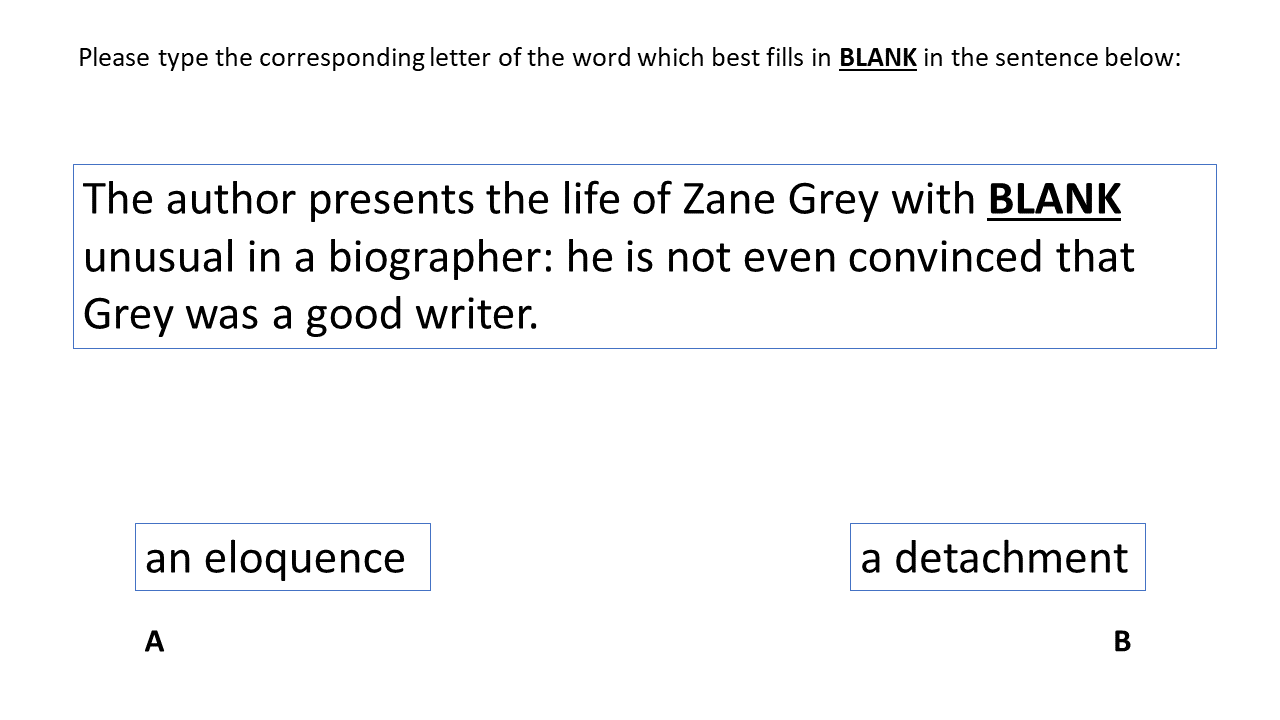}
    \caption{Reading: correct answer is ``B''.}
    \label{fig:readingStim}
\end{subfigure}
\hfill
\begin{subfigure}[b]{0.3\textwidth}
    \centering
    \includegraphics[width=\linewidth]{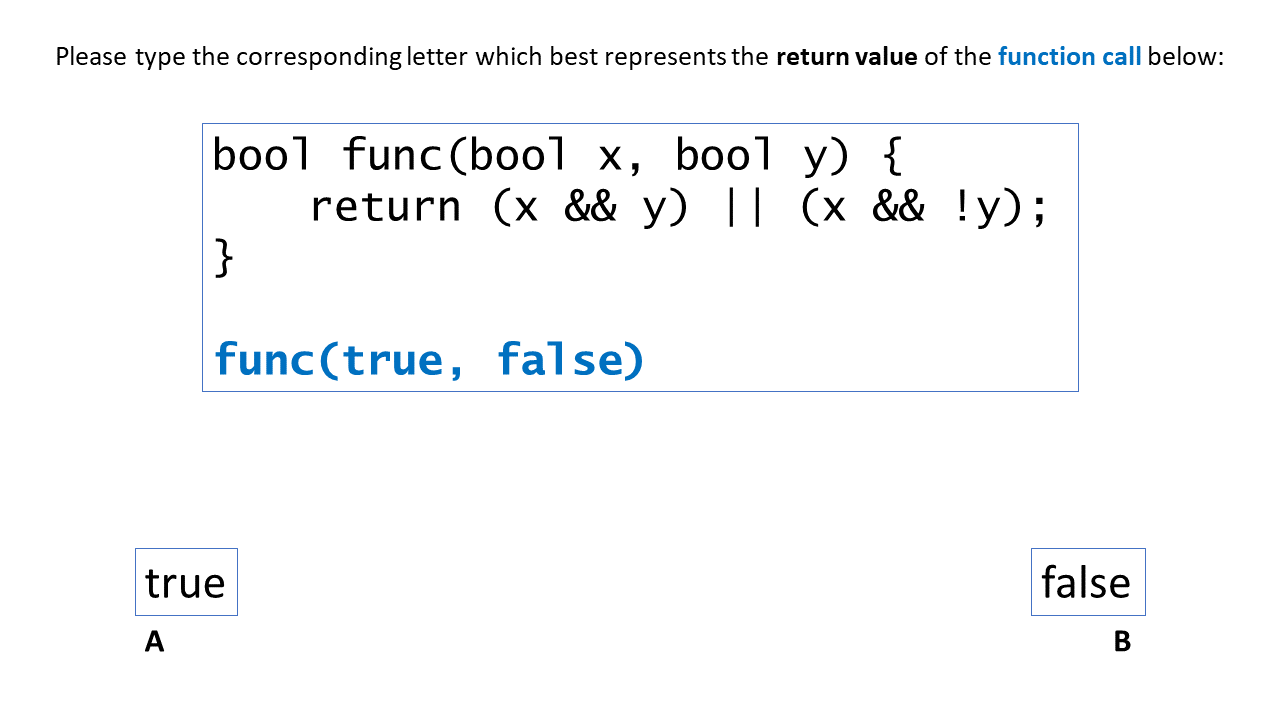}
    \caption{Code: correct answer is ``A''.}
    \label{fig:codingStim}
\end{subfigure}
\caption{Example fNIRS Stimuli for mental rotation $(a)$, reading $(b)$, and coding $(c)$ tasks.}

\end{figure*}

For the reading stimuli, we use sentence completion tasks adapted from official \emph{Graduate Record Examination} practice exam questions~\cite{GRE}, an assessment required for admission to many graduate programs. For each stimulus, participants were asked to read a sentence and choose the appropriate word or phrase to fill a blank (see Figure~\ref{fig:readingStim} for an example).

For the coding stimuli, we created a corpus of short code snippets that use constructs familiar to introductory computing students. Specifically, they contained boolean logic, while loops, for loops, and arrays. These are all early core concepts in most introductory curricula~\cite{tew2010developing} and are also covered in our institution's CS1. Unfortunately, we were not able to directly reuse coding stimuli from previous neuroimaging studies as they are generally geared toward expert programmers and thus contain constructs unfamiliar to novices. For the array-based questions, however, we were able to adapt stimuli created by Huang \emph{et al.}~\cite{huang19}. For each coding stimulus, participates were asked to choose either the correct output or return value of a short code snippet (see Figure~\ref{fig:codingStim} for an example).

\subsection{Followup Programming Assessment}
\label{sec:posttest}

At the end of the semester (10--12 weeks after the fNIRs scans), we invited participants to complete a written programming test. Along with the programming test, participants also completed a battery of cognitive, and behavioral assessments. For the programming assessment, we used the \emph{Second CS1 Assessment} (SCS1), a validated language-agnostic measure of CS1 programming ability~\cite{ParkerSCS1}. The SCS1 contains 27 multiple choice questions and takes one hour. It covers Boolean logic, while loops, for loops, arrays, if statements, functions, and recursion. There are three types of questions: definition questions, code tracing questions, and code replacement questions. Due to COVID-19, we were unable to hold the test in person. Rather, participants completed an online version of the SCS1 over a proctored video call. Responses were then checked for timing-based anomalies to ensure participants did not rush through the test. 

\section{Analysis Methodology}
\label{sec:analysis}

We now go over our methodology for analyzing the fNIRS data. Broadly, there are three stages in our analysis pipeline: preprocessing, individual modeling, and group level modeling.

\emph{1) Preprocessing:} The raw data, in the form of light intensity values, were converted into optical density data by calculating the fluctuations in light absorption by the presence of either oxygenated (HbO) or deoxygenated (HbR) blood. 
The optical density data were then converted into an HbO/HbR signal using the Modified Beer-Lambert law.
We ran a general linear model (GLM) with pre-whitening and robust least squares to fit the data~\cite{barker2013autoregressive}. 

\emph{2) Individual Subject Modeling:} After the hemodynamic response was modeled for each subject, quality control 
checks were implemented to limit the amount of noise in the group-level model. 
The signal-to-noise ratio, anticorrelation of HbO and HbR, and brain-activation plots were considered 
when deciding whether to exclude individual blocks or whole participants. 
The signal-to-noise ratio was calculated as a ratio between the absolute signal mean and standard deviation, with a
threshold set at 0.9. As a result, 16 blocks were excluded from further analysis (see Section~\ref{sec:fNIRSData_Collection} for a discussion of our experimental block setup).
In an ideal hemodynamic response function, the levels of HbO will increase as the levels
of HbR decrease and vice-versa~\cite{cui2010}, so the correlation between these two levels should be negative. 
Thirteen additional blocks that did not show this pattern were excluded from further analysis.

Next, the brain activations estimated by the GLM for \emph{All Conditions $>$ Rest} were plotted onto brain models
using a photogrammetry-based localization method~\cite{frank-photogram}, at which point the 
loci of activity could be examined and scrutinized.
We expected activity in the visual cortex (as this is a visual experiment), 
as well as in the inferior frontal gyrus during the reading task (as this is a well-documented region
for language processing).
Thirty seven blocks that did not pass this activation pattern check were excluded. These criteria were not mutually exclusive. In total, 59 blocks from 31 participants were included in the group-level modeling.

\emph{3) Group Modeling:} 
We used a linear mixed effects model for group level analysis, contrasting \emph{Task $>$ Baseline} activations
to estimate task-related brain activations and brain-behavior
correlations. Lastly, we applied a false-discovery rate (FDR) threshold
correction ($q < 0.05$) to account for the multiple-comparison issue.

\section{Validation}
\label{sec:validation}

In this section we validate that the brain activation patterns we observe during mental rotation and reading align with those established by previous work. Specifically, we present our results for the contrasts \emph{Mental Rotation $>$ Rest} and \emph{Reading $>$ Rest}. We provide brain activation visualizations for \emph{Mental Rotation $>$ Rest} and \emph{Reading $>$ Rest} in Figures~\ref{fig:rotationActivation} and~\ref{fig:readingActivation} respectively. We also provide a tabular view of our results with specific $t$-values in Table~\ref{tab:tValues} (see Section~\ref{sec:notation} for an overview of $t$-values and \emph{A $>$ B} notation). 

In \emph{Mental Rotation $>$ Rest}, we observe significant bilateral activation in the occipital and parietal lobes (BAs 7, 17, 18, 19, 39). The occipital lobe (BAs 17, 18, 19) is associated with the visual cortex and is responsible for tasks such as image and pattern recognition, both visiospatial processes. The parietal activation (BAs 7, 39) is also localized in regions associated with spatial tasks including spatial reasoning and mental rotation~\cite{culham2001neuroimaging, cohen1996changes}. 

Our results are therefore consistent with previous mental rotation neuroimaging; in his meta-review of mental rotation imaging studies, Zacks also identified BAs 7, 17, 19, and 39 as active during mental rotation tasks~\cite{zacks2008neuroimaging}. We do not observe significant supplementary motor cortex activation (BA 6), another area often active during mental rotation tasks. This is not too surprising, however, as the supplementary motor cortex is most strongly activated in tasks which encourage participants to use motor simulation strategies~\cite{zacks2008neuroimaging}; we did not prompt participants to use such strategies in our experiment.

\begin{figure}[t]
    \centering
    \includegraphics[width=0.8\linewidth]{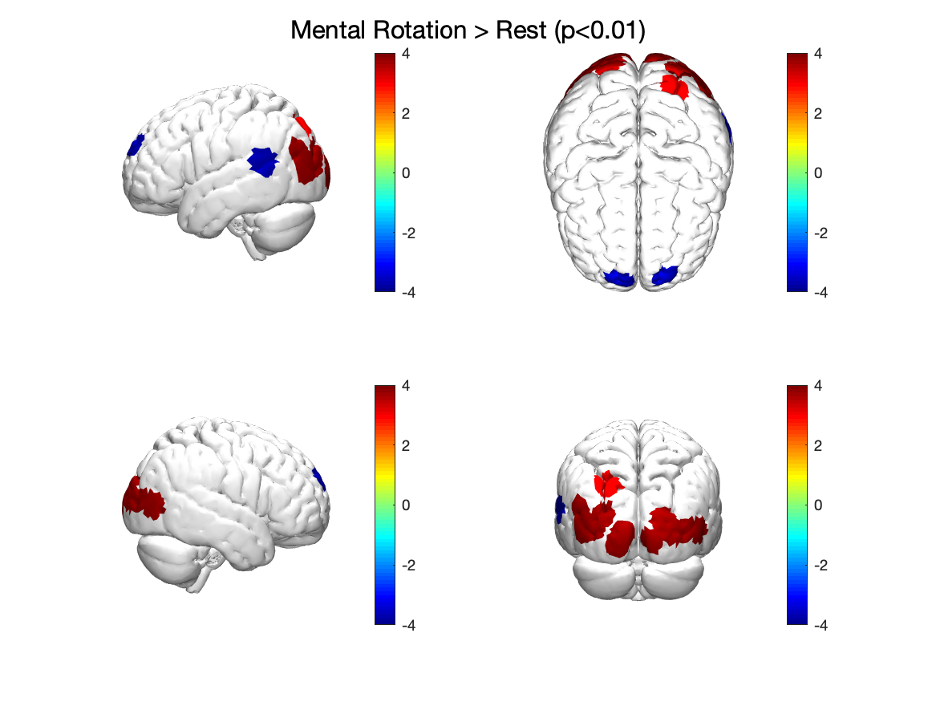}
    \caption{Baseline Mental Rotation Activation: Red indicates regions more activated during the mental rotation task while blue indicates regions more activated during rest. Note that all significant mental rotation activation is located in the back of the head in the parietal and occipital lobes.}
    \label{fig:rotationActivation}
\end{figure}

In \emph{Reading $>$ Rest}, we observe significant occipital, parietal, and prefrontal activation lateralized to the left hemisphere, aligning with previous work. We observe significant activation in both Broca's and Wernicke's areas, widely considered two of the most important language areas~\cite{price2012review}. We also observe significant activation in the left dorsolateral prefrontal cortex (See section ~\ref{sec:notation}, BA 46),
a region associated with attention and working memory. Regarding the occipital activation, while we observe some bilateral activation, significant activation is substantially more widespread in the left hemisphere. The left occipital cortex has been found to correspond with word and letter specific pattern recognition~\cite{cohen2003visual}.

\begin{figure}[t]
    \centering
    \includegraphics[width=0.8\linewidth]{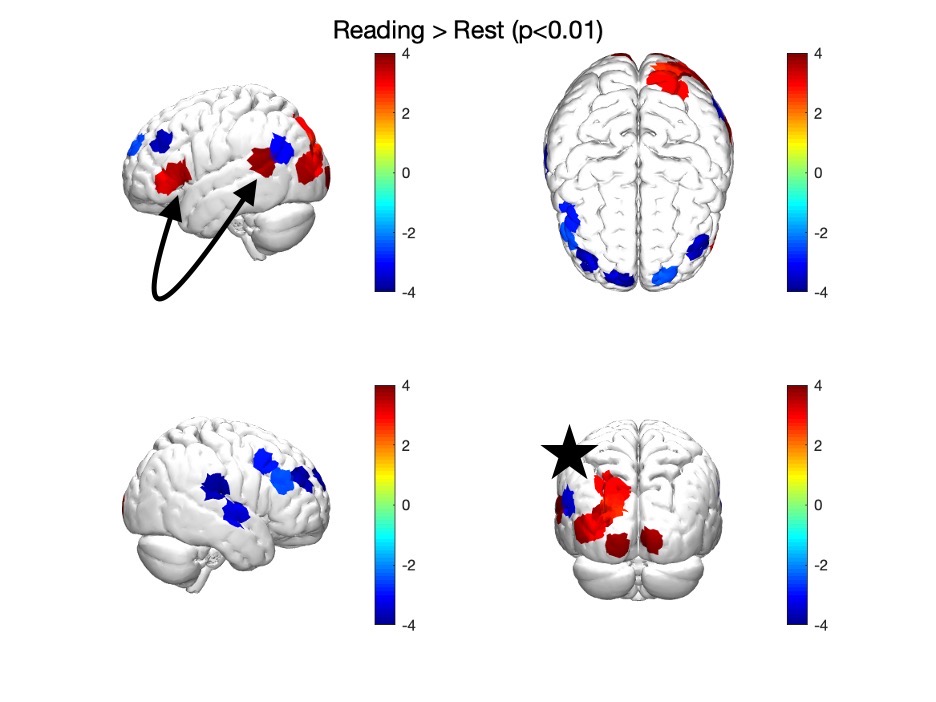}
    \caption{Baseline Reading Activation: Red indicates regions more activated during the reading task while blue indicates regions more activated during rest. Note the reading activation in Broca's and Wernicke's areas (the double arrow) as well as the left-lateralized occipital and parietal activation (the star).}
    \label{fig:readingActivation}
\end{figure}

\begin{framed}
    \noindent Our activations for rotation and reading align with previous work: significant occipital and parietal activation for rotation, and occipital and prefrontal activation for reading, including Broca's and Wernicke's areas. Rotation activation is bilateral while reading is primarily in the left hemisphere. This validation gives confidence for both construct and internal validity (i.e., that our protocol measures what we think it measures and does so correctly). 
\end{framed}

\section{Experimental Results}
\label{sec:results}

We now present the results of our experiment probing the neurological connections between reading, mental rotation, and programming for novice programmers (See Section~\ref{sec:notation} for an overview of relevant notation and vocabulary (e.g. \emph{A $>$ B}). We focus our results around three research questions:

\begin{itemize}
    \item RQ1---Programming Activation: What areas of the brain activate when novice software engineers program?
    \item RQ2---Comparative Activation: How does the coding brain activation of novices compare to their brain activation during mental rotation and during reading?
    \item RQ3---Prediction: Are there connections between coding brain activation patterns at the beginning of CS1 and their programming performance at the end of the course?
    \end{itemize}

\subsection{RQ1---Programming Activation} 

To determine which areas of the brain activate when novice software engineers program, we present the results of the contrast \emph{Code $>$ Rest}. That is, we test which brain areas significantly distinguish programming from a resting state ($p < 0.01$, $q < 0.05$). We also discuss the functionality of the distinguishing brain regions. Figure~\ref{fig:codingActivation} contains a brain activation visualization for \emph{Code $>$ Rest}, and we provide a tabular view of our results with specific $t$-values in Table~\ref{tab:tValues}. 

\begin{table*}
    \centering
    \begin{tabular}{lrrrrr}
        Brain Region & \emph{Rotation $>$ Rest} & \emph{Reading $>$ Rest} & \emph{Code $>$ Rest} & \emph{Code $>$ Rotation} & \emph{Code $>$ Reading} \\
        \bottomrule
        Frontal Cortex \\
        \toprule
        Left DLPFC (BA 46)  &   & 	$3.32-3.32$ &	$3.30-3.95$ &	$2.78-2.96$ &	$3.76-3.76$\\
        Right DLPFC (BA 46) &   & 	$-2.62--2.62$ &	$2.78-3.27$ &	$3.79-3.79$ &	$2.99-4.72$\\
        Broca's Area (Left BAs 44 and 45)  & & 	$3.76-3.76$	 & & 	 & 	$-3.54--3.54$\\
        IFG (Right BAs 44 and 45)  & & 	 & 	$2.78-3.27$&	$3.11-3.79$&	$2.99-2.99$\\	
        Left Supplementary Motor Cortex (BA 6)  & & 	  & 	$3.45-3.45$&	$3.45-3.45$&	\cellcolor{blue!25}$3.22-5.32$\\ 
        Right Premotor Cortex (BA 6)  & & 	$-2.98--2.98$	 & & 	 & 	  \\ 
        Left BA 8  & & 	$-4.07--4.07$&	$3.30-3.30$&	$2.96-2.96$&	$-4.93-3.93$ \\
        Right BA 8  & & 	$-3.95--2.62$	 & & 	 &	$2.67-4.72$ \\
        Left BA 9 & \cellcolor{blue!25}$-6.26--6.26$&	$-4.07--2.63$&	\cellcolor{blue!25}$-7.49--7.49$	 & & 	\cellcolor{blue!25}$-5.10-3.93$ \\
        Right BA 9 &\cellcolor{blue!25}$-7.46--7.46$&	\cellcolor{blue!25}$-6.69--2.67$&	\cellcolor{blue!25}$-6.13--6.13$	 & & 	$2.67-4.72$\\
        \bottomrule
        Temporal Cortex \\
        \toprule 
        Wernicke's Area (Left BAs 22, 40) & \cellcolor{blue!25}$-6.44--6.44$&	\cellcolor{blue!25}$5.91-5.91$&	&	\cellcolor{blue!25}$6.51-6.51$&	$-4.77--4.77$\\
        Right BA 21  & & 	 & 	 &	 &	$2.79-2.79$ \\
        Right BA 22  & & 	$-3.65--3.65$	 & & 	 &	$3.15-3.15$\\
        Right Auditory Cortex (BA 41)  & & 	\cellcolor{blue!25}$-5.81--5.81$	 & & 	 &	\cellcolor{blue!25}$6.70-6.70$ \\
        \bottomrule 
        Parietal Cortex \\
        \toprule 
        Left BA 7 & $3.38-3.38$&	$3.15-3.17$&	$3.64-3.64$&	$2.60-2.60$	 & \\
        Left Angular Gyrus (BA 39) &$4.95-4.95$&	$-3.19--3.19$&	\cellcolor{blue!25}$-2.99-5.28$	 & & \cellcolor{blue!25}$6.86-6.86$\\
        Right Angular Gyrus (BA 39) & \cellcolor{blue!25}$6.83-6.83$	 & &\cellcolor{blue!25}$3.68-5.15$&	$-3.49-3.98$&	$4.29-4.29$\\
        \bottomrule 
        Occipital Cortex \\
        \toprule 
        Left BA 17 & \cellcolor{blue!25}$5.78-6.36$&\cellcolor{blue!25}$8.11-8.11$&	$2.75-2.83$&	$-3.26--3.23$&	$-4.48-4.78$\\
        Right BA 17 & $4.17-4.17$&	$4.29-4.29$	 & & 	$-2.84--2.84$ &	$-2.91--2.91$\\
        Left BA 18 & \cellcolor{blue!25}$4.69-6.36$&\cellcolor{blue!25}$3.06-8.11$&	$2.75-3.84$&	$-3.23--3.23$&	$-4.48--4.48$ \\
        Right BA 18 & \cellcolor{blue!25}$4.18-6.48$ &	$4.29-4.29$&\cellcolor{blue!25}$5.95-5.91$ &	$-2.84--2.84$&	\cellcolor{blue!25}$-2.91-5.26$ \\
        Left BA 19 & \cellcolor{blue!25}$4.52-7.76$ &	$2.71-4.16$ &\cellcolor{blue!25}$3.84-5.46$	 &  &	 \\
        Right BA 19 & \cellcolor{blue!25}$3.95-6.83$	 & & \cellcolor{blue!25}$3.68-5.95$ &	$-3.49--3.49$ &	\cellcolor{blue!25}$3.54-5.65$\\
        \toprule

    \end{tabular}
    \caption{$t$-value statistics for \emph{Mental Rotation $>$ Rest}, \emph{Reading $>$ Rest}, \emph{Code $>$ Rest}, \emph{Code $>$ Mental Rotation}, and \emph{Code $>$ Reading}. All reported results are significant ($p < 0.01$) and pass our false discovery threshold ($q < 0.05$). Blank cells indicate no significant effect was found in that region. Results closer to +8 or -8 indicate more significant activation or deactivation. We highlight results with t-values less than -5 or greater than +5.}
    \label{tab:tValues}
\end{table*}

While coding, novices exhibit significant occipital activation (BAs 17, 18, 19). While bilateral, we observe somewhat stronger right hemisphere activity. Functionally, the occipital cortex is associated with visual processing, and it includes areas such as the primary visual cortex (BA 17) and visual association area (BA 18). 
This occipital activation is the strongest activation we observe in \emph{Code $>$ Rest}, with three out of the five channels with $t$-values greater than five. 

\begin{figure}[t]
    \centering
    \includegraphics[width=0.8\linewidth]{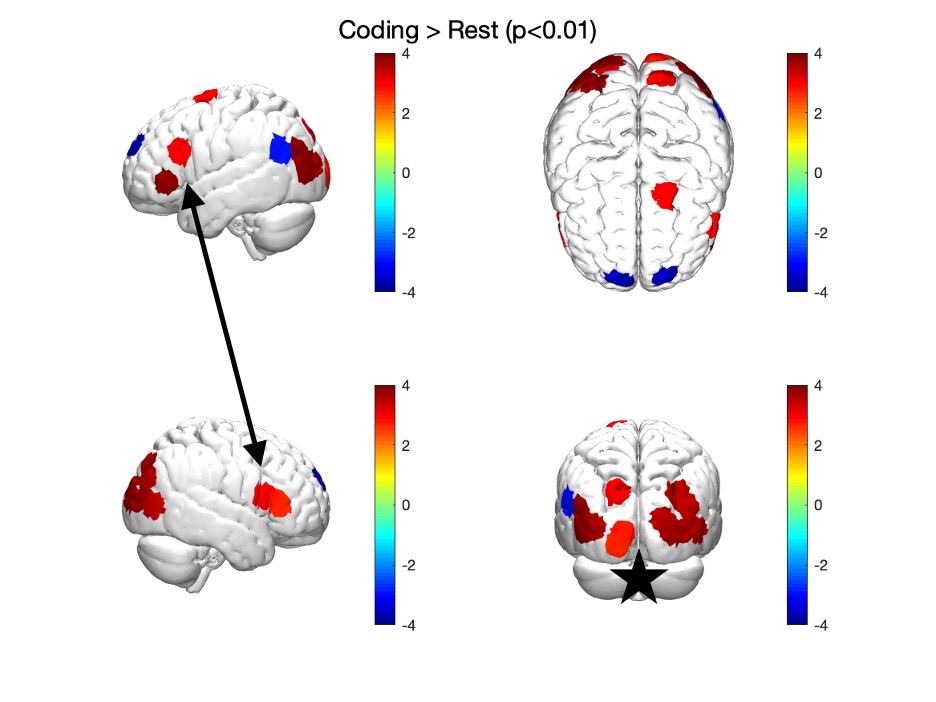}
    \caption{Significant Coding Activation: Red indicates regions more activated during the coding task while blue indicates regions more activated during rest. Note the widespread bilateral activation in both the DLPFC (the arrows) and the occipital and posterior parietal cortices (the star).}
    \label{fig:codingActivation}
\end{figure}

Beyond occipital activation, we also observe significant posterior parietal activation, primarily in the angular gyrus (BA 39). This activation is bilateral: the other two channels with $t$-values greater than five both cover BA 39, one in each hemisphere. 
In the left hemisphere, the angular gyrus is important for language-related tasks~\cite{brownsett2010contribution}. Some researchers also include BA 39 in Wernicke's area, one of the two main brain cortex regions associated with natural language processing. However, we do not observe activation while coding in the regions most commonly associated with Wernicke's area: BAs 22 and 40. 
The angular gyrus is also strongly associated with spatial cognition tasks including spatial orientation (e.g., distinguishing left from right), spatial attention, numerical computation, and mental rotation~\cite{seghier2013angular, zacks2008neuroimaging}. 
While the spatial functionality of the angular gyrus is bilateral, many spatial tasks, including mental rotation, are concentrated in the right hemisphere~\cite{seghier2013angular}. Therefore, the bilateral activation of the angular gyrus indicates that \emph{novices use both language and spatial cognitive processes while programming}.

We also observe significant activation in the frontal cortex. Specifically, we observe bilateral activation in the DLPFC (BA 46) and activation in the left superior premotor cortex (BA 6). The premotor cortex is associated with motor processing, and it has also been found to activate during visiospatial tasks including mental rotation~\cite{zacks2008neuroimaging}. The DLPFC is associated with working memory; lower activation in this region corresponds with worse performance on working-memory intensive tasks such as complex problem solving~\cite{barbey2013dorsolateral}. The significant bilateral DLPFC activation, therefore, indicates that \emph{novices find programming a challenging and working memory intensive task}.  

\begin{framed}
    \noindent 
    Novices engage brain regions associated with language and spatial cognition, as well as regions associated with increased demand for attention and executive function (via neural activity in \emph{Code $>$ Rest}, $p < 0.01$, $q < 0.05$).

\end{framed} 

\subsection{RQ2---Significant Comparative Activation}

We now compare novices' coding brain activation to their activation during mental rotation and reading by presenting our findings for the contrasts \emph{Code $>$ Mental Rotation} and \emph{Code $>$ Reading}. 
We provide brain activation visualizations for \emph{Mental Rotation $>$ Rest} and \emph{Reading $>$ Rest} in Figures~\ref{fig:codingRotationActivation} and~\ref{fig:codingReadingActivation}, and we provide a tabular view of our results with specific $t$-values in Table~\ref{tab:tValues}. 
Our high level finding is that while coding, mental rotation, and reading are all neurally distinct tasks, \emph{we observe more substantial differences between coding and reading than we do between coding and mental rotation}.

\begin{figure}[t]
    \centering
    \includegraphics[width=0.8\linewidth]{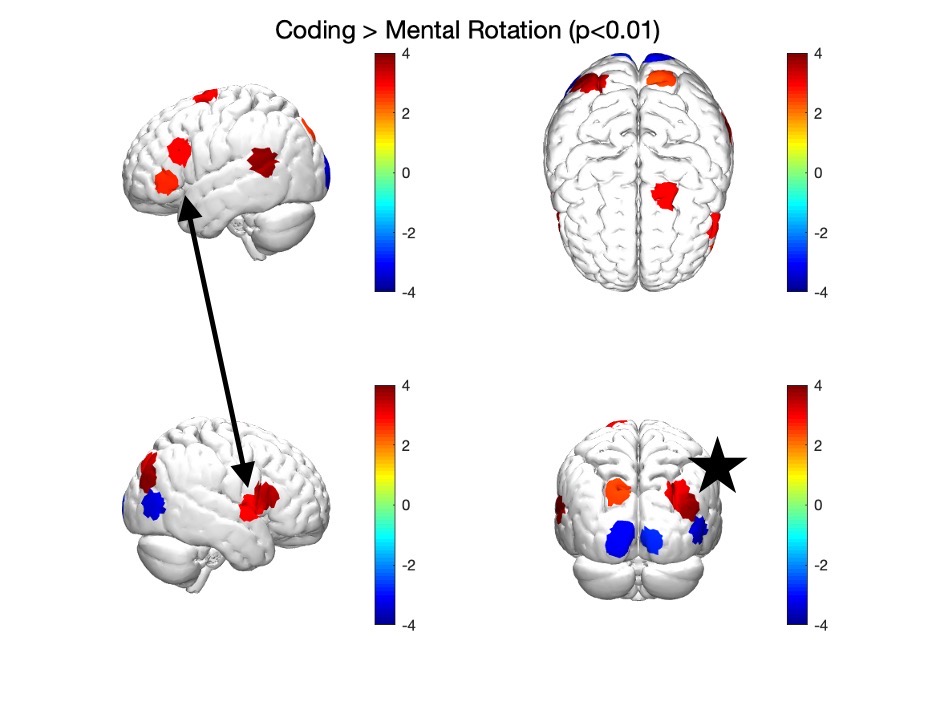}
    \caption{Activation Contrast between Programming and Mental Rotation: Red indicates regions more activated during coding while blue indicates regions more activated during mental rotation. Note that compared to mental rotation, coding has stronger bilateral frontal activation (the arrow) and right posterior parietal activation in the angular gyrus (the star).}
    \label{fig:codingRotationActivation}
\end{figure}

\begin{figure}[t]
    \centering
    \includegraphics[width=0.8\linewidth]{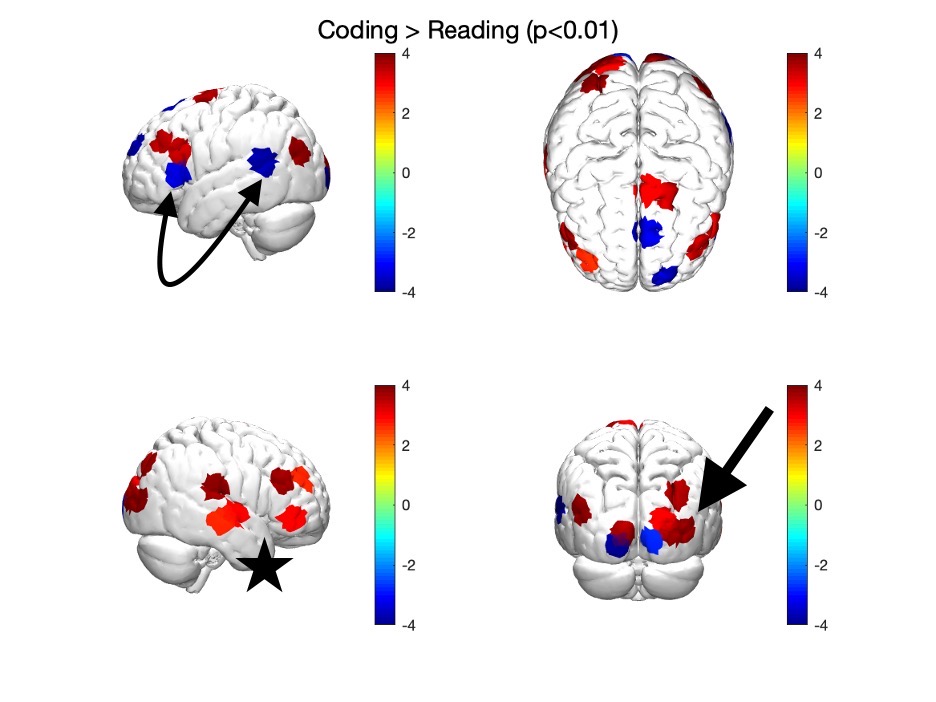}
    \caption{Activation Contrast between Programming and Reading: Red indicates regions more activated during coding while blue indicates regions more activated during reading. Note that reading has comparatively more activation in Broca's and Wernicke's areas (the double arrow), while coding has substantially more right frontal activation (the star) and more right-lateralized occipital / parietal activation (the single arrow).}
    \label{fig:codingReadingActivation}
\end{figure}

For \emph{Coding $>$ Mental Rotation}, we observe several significant differences in activation. First, when coding, participants exhibit more bilateral frontal activation than while mentally rotating objects. This comparative activation is in the DLPFC and the premotor cortex (BAs 6 and 46), areas commonly associated with working memory and spatial manipulation~\cite{barbey2013dorsolateral, zacks2008neuroimaging}. Similarly, we observe comparatively high coding activation in the right angular gyrus, a region also connected with spatial reasoning~\cite{seghier2013angular}. We find it intriguing that many of the areas with more activity for coding than for mental rotation are associated with spatial reasoning, ostensibly the process measured by the mental rotation stimuli. This may imply that coding is a challenging task that is actually \emph{more} spatially intensive for novices than simple mental rotation. 

When considering a contrast \emph{A $>$ B}, 
it is important to distinguish between a comparison that is positive because $A$ is greater than $B$ (called a ``true activation'') 
vs. one that is positive only because $B$ is negative (a ``strong deactivation''). 
All of the significant \emph{Coding $>$ Mental Rotation} differences discussed so far are channels with either positive activation or no significant activation in \emph{Rotation $>$ Rest}. Thus, they are all true differences between coding and mental rotation. On the other hand, while we also observe a very strong comparative activation ($t=6.51$) in \emph{Coding $>$ Mental Rotation} in Wernicke's area, this is a facet of the same channel's strong deactivation in \emph{Rotation $>$ Rest} instead of a true activation.

For \emph{Coding $>$ Reading}, we also observe significant differences in activation: coding has comparatively stronger activation throughout the right hemisphere and in the premotor cortex while reading has stronger activation in Broca's and Wernicke's areas. All of the significant differences in the left hemisphere including Broca's and Wernicke's areas are true contrasts; that is, none of them are caused by a significant deactivation in one of the \emph{Task $>$ Rest} comparisons. In the right hemisphere, several of the apparent activations for coding are caused by strong deactivation in \emph{Reading $>$ Rest}. Even so, we observe true comparative right-hemisphere activations: the right occipital, angular gyrus, and DLPFC are all more activated while coding than while reading. 

While we observe that Coding is neurologically distinct from both mental rotation and reading, we also observe more substantial differences in \emph{Coding $>$ Reading} than in \emph{Coding $>$ Mental Rotation}. In \emph{Coding $>$ Reading}, there are four channels with $t$-values greater than $+5$ or less than $-5$, while in \emph{Coding $>$ Mental Rotation}, there is only one (see Table~\ref{tab:tValues}).  Furthermore, the strong channel difference in \emph{Coding $>$ Mental Rotation} is not particularly compelling as it is caused by deactivation. Taken together with the previous functional analysis, this trend hints that for novices, coding is a more spatially based and a less language-based cognitive process. 

\begin{framed}

    \noindent We find that, for novices, coding is neurally distinct from both reading and spatial reasoning. 
    Coding engages regions associated with working memory
    more than does either reading or rotation, indicating that programming is a more cognitively challenging task. However, we observe more significant substantial differences in \emph{Coding $>$ Reading} than we do in \emph{Coding $>$ Rotation}: novices may rely heavily on spatial reasoning while coding.
\end{framed} 

\subsection{RQ3---Prediction}

We now turn to an exploratory analysis of connections between observed brain activation patterns and participants' final programming assessment scores. To do so, 
we use \emph{Representational Similarity Analysis} (RSA), a common Psychology approach, to correlate brain activity interactions with scores on the programming post-test (see Kriegeskorte \emph{et al.}~\cite{kriegeskorte2008representational} for an introduction to RSA). We test if the brain activation similarity between mental rotation and coding (\emph{Mental Rotation $\times$ Coding}) or the brain activation similarity between reading and coding (\emph{Reading $\times$ Coding}) are correlated with SCS1 scores. We calculate these correlations for the right hemisphere, left hemisphere, right hemisphere frontal, left hemisphere frontal, and occipital regions for a total of five statistical tests per hypothesis and ten total test--hypothesis pairs.

We find a correlation that remains significant after applying the Bonferroni correction for multiple comparisons~\cite{chen2017general}. Specifically, we find a significant medium negative correlation between (\emph{Mental Rotation $\times$ Coding}) in the right frontal region and programming post-test score ($r = -0.48$, $p=0.0006$, adjusted Bonferroni $p=0.006$): the \emph{less similar} the neural activation patterns for coding and rotation, the \emph{better} the final programming assessment outcome.

Notice that coding also elicited stronger right frontal activation than the mental rotation task. It is therefore possible that the more individuals engage the right frontal, in a way that is dissimilar from its engagement during spatial processing, the less progress they make over the course of the semester. 
The hypothesis is similar to the observation that the more dissimilar is the activity between novice readers and their language, the less progress they make in learning to read~\cite{marks-kovelman-2019}.
In the Psychology literature, one common approach for testing such a hypothesis would be to investigate correlations between significant channel activations and response time. 
However, while in some settings response time can be used as a proxy for difficulty, our coding stimuli
were not designed with that consideration; we do not find a statistically significant correlation ($r=-0.095$, $p=0.4866$) and lack enough information to either substantiate or refute that hypothesis. 


\begin{framed}
    \noindent In an exploratory analysis relating initial brain activation patterns and post-test programming scores, 
    we find that \emph{less similar} patterns of activation for coding and mental rotation in the right
    frontal hemisphere at the start of the semester predict better outcomes on the end-of-semester final programming assessment ($r = -0.482$, $p=0.006$).
\end{framed}

\section{Discussion}

In this section, we discuss the implications of our experimental results. In particular, we consider how the coding brain activation patterns of novice programmers compare to those of more experienced software developers (Section~\ref{sec:novicesVexperts}) and also discuss future research directions (Section~\ref{sec:future}). 

\label{sec:discussion}

\subsection{Novices vs. Experts}
\label{sec:novicesVexperts}

In this section, we discuss how our results compare to those observed in neuroimaging studies of expert developers. Generally, we observe more right-hemisphere activation, more engagement of visiospatial processes, and less engagement of language processes than is seen in experts. For example, all significant activation areas observed by Siegmund~\emph{et al.} were in the left hemisphere, a majority coinciding with established language regions~\cite{siegmund2014understanding}. Similarly, Floyd~\emph{et al.} found that for experts, programming becomes increasingly less distinguishable from reading, a left-lateralized cognitive activity. In this context, our work helps establish an  experience-based lateralization shift: novice programmers generally exhibit bilateral activation, especially in regions associated with visiospatial processing, while expert developers see increasingly left-lateralized activation centered in language-associated regions.

However, not all extant studies of experts observe a strong connection between reading and programming: in their study examining \emph{code writing} (as opposed code reading, the focus of other neuroimaging studies including our own), Krueger \emph{et al.} observed significantly more right-brain activation in spatial areas during code writing than prose writing~\cite{krueger2020}. More investigation is needed to see if code writing exhibits a similar lateralization trend with expertise. However, it is possible that code writing consistently remains a more spatial activity.

\subsection{Future Directions and Implications}
\label{sec:future}

To the best of our knowledge, this is the first paper to use medical imaging to explicitly investigate 
novice programmer coding brain activation patterns and their correlates. As a result, beyond replications and meta-analyses, which
are more common in Psychology (e.g.,~\cite{ventura2005identifying}) but not yet as prevalent in Computer Science,
many of the research implications relate to building on the baselines established by the results presented here.
We focus on two dimensions: how programmers learn other computing activities at a cognitive level, and how learning
programming in general compares to established neurological theories for learning other disciplines. 

For the former, we note that recent medical imaging research on programming has focused on program comprehension
and code review, with a lesser emphasis on data structures and code writing~\cite{siegmund2014understanding, FSE17, huang19, floyd2017decoding, krueger2020}. Other activities
remain unexplored. To take one example, it is unknown whether boolean logic has a significant spatial cognitive
component. While general logic has been studied, particular paradigms, such as circuit design, may be processed differently by humans, 
potentially suggesting alternate training or tool-support approaches. Other experimental protocol paradigms
are also relatively unexplored: almost all software engineering neuroimaging research consists of showing
fixed, static stimuli. Even relatively foundational experimental structures in Psychology, such as priming, masking, and recall are unexplored. For example, building on the influential work of Chase and
Simon~\cite{simon1973chess}, Psychologists have studied the relationship between chess expertise and the ability to 
recall or reason about briefly-presented random chess boards~\cite{gong2015chess}. While Siegmund
\emph{et al.} randomized aspects of programs to study comprehension~\cite{FSE17}, using such paradigms to tease apart programming expertise neurologically remains unexplored. 

For the latter, we observe that a number of theories and hypotheses about how humans learn various subjects, from 
second languages~\cite{kovelman2008} to musical instruments~\cite{lathorp1970}, have been posited in the literature. 
We believe that it will be fruitful to investigate whether a sequential model or a more spatial encoding strategy (see Margulieux~\cite{MargulieuxSpatialTheory}) best describes learning to program. 
Based on our results, our preliminary speculation is that spatial encoding is indeed a key general strategy employed by novices that may decrease in importance over time. If true, this would have implications for the use of domain-specific strategies in skills-based training. A more concrete investigation for programming is merited. 

\section{Limitations and Threats to Validity}
\label{sec:threats}

Although our experiments and analysis
provide significant evidence about novice
programmers and spatial ability, our 
results may not generalize. We consider
a number of threats to validity and discuss 
how our approach mitigates them.

We note that fNIRS experiments are dependent on transmitters and receivers for infrared light (see Section~\ref{sec:medicalImaging}): if no pairs
are present for a relevant portion of the brain, activity there cannot be measured. We mitigate
this potential source of false negatives in two ways. First, adapt a validated cap design proposed by Huang \emph{et al.} for use in
software engineering and spatial ability
studies~\cite{huang19}. Second, information
from other medical imaging approaches, such as fMRI, which do not depend on transmitter placement, is used to determine which brain regions to measure~\cite{siegmund2014understanding,floyd2017decoding,krueger2020}. 

We also consider issues of construct
validity: are we measuring what we claim 
to be measuring (e.g., spatial ability,
introductory programming, etc.)? 
While there are multiple aspects to
spatial ability, we 
use mental rotation,
an established paradigm
for investigating spatial ability, both in
psychology in general~\cite{harris2000selective,culham2001neuroimaging,cohen1996changes}
and in computer science in particular~\cite{bockmon2020cs1, huang19}. For introductory programming, we make use of the
SCS1, a validated assessment~\cite{ParkerSCS1}. 

Finally, all of our subjects are students at the 
same large US university. This aspect of 
participation selection may limit the
generality of our results to other populations. 

\section{Related Work}
\label{sec:related}

We place our results in context with respect to three broad categories of previous work. 

\paragraph{Spatial Skills and Programming} 
There is a positive relationship between programming and spatial ability and~\cite{jones2008spatial, fincher2006predictors, ParkerSocio, parkinson2018investigating}. Parkinson and Cutts found that ``spatial skills typically increase as the level of academic achievement in computer science increases''~\cite{parkinson2018investigating}. Furthermore, Parker \emph{et al.} found that spatial reasoning is better mediating variable for affluence discrepancies in computer science than computing access~\cite{ParkerSocio}. 
There have also recently been studies establishing a causal transfer between spatial reasoning training and computer science performance. Cooper \emph{et al.} and Bockmon \emph{et al.} ran studies with high school and university programmers, finding that those who participated in additional spatial training performed better on a final programming test~\cite{cooper15,bockmon2020cs1}. 
Margulieux's spatial encoding strategy (SpES) framework relates the cogitative processes behind spatial ability and learning to program~\cite{MargulieuxSpatialTheory}. SpES hypothesizes that strong spatial reasoning ability helps novice programmers use general strategies for mentally encoding non-verbal information.

Our results provide context and nuance to such claims: we find that
novice programmers \emph{do} use spatial cognitive processes while programming (RQ1) and that the degree of dissimilarity between patterns of neural activity for coding and spatial tasks \emph{can} predict final outcomes (RQ3). 

\paragraph{Reading and Programming}
From documentation to code review to requirements elicitation to code summarization, many software engineering activities involve
a significant reading component~\cite{xia2017measuring,haiduc2010use,mcburney2015automatic}. 
Experimentally, several studies report a correlation between overall programming ability and the ability to read a program and describe its function in natural language~\cite{murphy2012ability, lopez2008relationships} or posit
natural language reading as a basis for code comprehension~\cite{busjahn2011analysis, prat2020relating}. Some models extend to
training: Fedorenko \emph{et al.} hypothesize that 
``pedagogies for developing linguistic fluency'' can inform how to train programmers, based on a perceived similarity
between learning programming and second language learning~\cite{fedorenko2019language}. Their hypothesis was supported by a recent study by Prat \emph{et al.} which found that natural language aptitude was a significant factor in predicting programming success~\cite{prat2020relating}. 

Our results elaborate on how such claims apply to novice programmers: while novices \emph{do} use language cognitive processes when coding (RQ1), we find that coding and reading are more \emph{different} for novices than are coding and spatial ability, suggesting they rely more on spatial reasoning when coding. 

\paragraph{Medical Imaging and Software Engineering}
Following the pioneering work of Siegmund \emph{et al.}~\cite{siegmund2014understanding}, a
number of papers have used medical imaging techniques to investigate software
engineering activities (e.g.,~\cite{floyd2017decoding, krueger2020, nakagawa2014quantifying, ikutani2014brain, fakhoury2018effect, Duraes16, Castelhano2018, siegmund2014understanding, FSE17}). Related to our work in particular, Yu \emph{et al.} used fNIRS to compare mental rotation tasks to data structure manipulation~\cite{huang19}.

A key distinction of our work is that those studies focus on programmers with years of experience. Explicit
investigations of programming expertise using neuroimaging are relatively rare, and tend to involve either
proxies such as undergraduate grades~\cite{floyd2017decoding} or comparisons between graduates or professionals
and undergraduates (e.g., Siegmund \emph{et al.} measure 8 students and 3 professionals~\cite[Sec.~3.3]{FSE17}). 
While Floyd \emph{et al.} found that coding and prose tasks are more similar
in terms of neural activity for senior undergraduate than for mid-level undergraduates~\cite{floyd2017decoding} (i.e., as
programmers become more experienced), our results provide evidence that \emph{the pattern continues}: as programmers become less experienced, programming and reading show less cognitive similarity (RQ1, RQ2). 
\section{Conclusion}

Neurological understandings of how novices engineer software has implications for training, pedagogy and
tool development. In a study of 31 participants, we use fNIRS to 
compare the neural activity for introductory programming, reading and spatial reasoning tasks in a 
controlled, contrast-based experiment.
We find that \textbf{all three tasks --- coding, 
prose reading, and mental rotation --- are mentally distinct
for novices}. This clarifies previous findings that they may be more similar
in experts~\cite{floyd2017decoding} or for complex data structures~\cite{huang2019distilling}. 
However, while those tasks are neurally distinct, we find 
\textbf{more significant and substantial differences between prose and
coding than between mental rotation and coding}. 
Intriguingly, we find generally
more activation in areas of the brain associated with spatial ability 
and task difficulty while coding compared to that reported studies with
more expert developers.
Finally, in an exploratory analysis, we find that \textbf{certain patterns
of neural activity at the start of the semester are predictive of
end-of-semester outcomes}, opening the door for future experiments
to model such phenomena more directly. 
To the best of our knowledge, this is the first study to focus specifically
on novice programmers and to make use of a significant time-delayed outcomes
assessment.
While preliminary, these findings both elaborate on
previous results (e.g., relating expertise to a similarity
between coding and prose reading) and also provide a
new understanding the cognitive processes underlying
novice programming.

\section*{Acknowledgements}

We acknowledge the partial support of the NSF (CCF
1908633, CCF 1763674) as well as both the Center for Research on Learning and Teaching and also the Center for Academic Innovation at the University of Michigan. Additionally, we thank Jessica Kim for her help understanding the fNIRS setup, and we thank Yu Huang for sharing the fNIRS cap from her previous work. Finally, we thank our undergraduate research assistants Anne Fitzpatrick, Annie Li, and Serena Chan for their logistical help and their help piloting fNIRS stimuli.


\bibliographystyle{IEEEtran}
\bibliography{IEEEabrv,sample-base}

\begin{thebibliography}{10}
\providecommand{\url}[1]{#1}
\csname url@samestyle\endcsname
\providecommand{\newblock}{\relax}
\providecommand{\bibinfo}[2]{#2}
\providecommand{\BIBentrySTDinterwordspacing}{\spaceskip=0pt\relax}
\providecommand{\BIBentryALTinterwordstretchfactor}{4}
\providecommand{\BIBentryALTinterwordspacing}{\spaceskip=\fontdimen2\font plus
\BIBentryALTinterwordstretchfactor\fontdimen3\font minus
  \fontdimen4\font\relax}
\providecommand{\BIBforeignlanguage}[2]{{%
\expandafter\ifx\csname l@#1\endcsname\relax
\typeout{** WARNING: IEEEtran.bst: No hyphenation pattern has been}%
\typeout{** loaded for the language `#1'. Using the pattern for}%
\typeout{** the default language instead.}%
\else
\language=\csname l@#1\endcsname
\fi
#2}}
\providecommand{\BIBdecl}{\relax}
\BIBdecl

\bibitem{siegmund2014understanding}
J.~Siegmund, C.~K{\"a}stner, S.~Apel, C.~Parnin, A.~Bethmann, T.~Leich,
  G.~Saake, and A.~Brechmann, ``Understanding understanding source code with
  functional magnetic resonance imaging,'' in \emph{Proceedings of the 36th
  International Conference on Software Engineering}, 2014, pp. 378--389.

\bibitem{FSE17}
\BIBentryALTinterwordspacing
J.~Siegmund, N.~Peitek, C.~Parnin, S.~Apel, J.~Hofmeister, C.~K\"{a}stner,
  A.~Begel, A.~Bethmann, and A.~Brechmann, ``{Measuring Neural Efficiency of
  Program Comprehension},'' in \emph{Foundations of Software Engineering},
  2017, pp. 140--150. [Online]. Available:
  \url{http://doi.acm.org/10.1145/3106237.3106268}
\BIBentrySTDinterwordspacing

\bibitem{floyd2017decoding}
B.~Floyd, T.~Santander, and W.~Weimer, ``Decoding the representation of code in
  the brain: An fmri study of code review and expertise,'' in \emph{2017
  IEEE/ACM 39th International Conference on Software Engineering (ICSE)}.\hskip
  1em plus 0.5em minus 0.4em\relax IEEE, 2017, pp. 175--186.

\bibitem{nakagawa2014quantifying}
T.~Nakagawa, Y.~Kamei, H.~Uwano, A.~Monden, K.~Matsumoto, and D.~M. German,
  ``Quantifying programmers' mental workload during program comprehension based
  on cerebral blood flow measurement: a controlled experiment,'' in
  \emph{Companion proceedings of the 36th international conference on software
  engineering}, 2014, pp. 448--451.

\bibitem{fakhoury2018effect}
S.~Fakhoury, Y.~Ma, V.~Arnaoudova, and O.~Adesope, ``The effect of poor source
  code lexicon and readability on developers' cognitive load,'' in
  \emph{International Conference on Program Comprehension}, 2018.

\bibitem{huang19}
\BIBentryALTinterwordspacing
Y.~Huang, X.~Liu, R.~Krueger, T.~Santander, X.~Hu, K.~Leach, and W.~Weimer,
  ``Distilling neural representations of data structure manipulation using
  {fMRI} and {fNIRS},'' in \emph{International Conference on Software
  Engineering}, 2019, pp. 396--407. [Online]. Available:
  \url{https://doi.org/10.1109/ICSE.2019.00053}
\BIBentrySTDinterwordspacing

\bibitem{krueger2020}
R.~Krueger, Y.~Huang, X.~Liu, T.~Santander, W.~Weimer, and K.~Leach,
  ``Neurological divide: An fmri study of prose and code writing,'' in
  \emph{International Conference on Software Engineering}, 2020.

\bibitem{buxton2004modeling}
R.~B. Buxton, K.~Uluda{\u{g}}, D.~J. Dubowitz, and T.~T. Liu, ``Modeling the
  hemodynamic response to brain activation,'' \emph{Neuroimage}, vol.~23, pp.
  S220--S233, 2004.

\bibitem{boas2014twenty}
D.~A. Boas, C.~E. Elwell, M.~Ferrari, and G.~Taga, ``Twenty years of functional
  near-infrared spectroscopy: introduction for the special issue,'' 2014.

\bibitem{lloyd2010illuminating}
S.~Lloyd-Fox, A.~Blasi, and C.~Elwell, ``Illuminating the developing brain: the
  past, present and future of functional near infrared spectroscopy,''
  \emph{Neuroscience \& Biobehavioral Reviews}, vol.~34, no.~3, pp. 269--284,
  2010.

\bibitem{ehlis2014application}
A.-C. Ehlis, S.~Schneider, T.~Dresler, and A.~J. Fallgatter, ``Application of
  functional near-infrared spectroscopy in psychiatry,'' \emph{Neuroimage},
  vol.~85, pp. 478--488, 2014.

\bibitem{obrig2014nirs}
H.~Obrig, ``{NIRS} in clinical neurology --- a `promising' tool?''
  \emph{Neuroimage}, vol.~85, pp. 535--546, 2014.

\bibitem{scicurious_2012}
\BIBentryALTinterwordspacing
Scicurious, ``{IgNobel} prize in neuroscience: The dead salmon study,''
  \emph{Scientific American Blog Network}, Sep 2012. [Online]. Available:
  \url{https://blogs.scientificamerican.com/scicurious-brain/ignobel-prize-in-neuroscience-the-dead-salmon-study/}
\BIBentrySTDinterwordspacing

\bibitem{bennett2009neural}
C.~M. Bennett, M.~Miller, and G.~Wolford, ``Neural correlates of interspecies
  perspective taking in the post-mortem atlantic salmon: an argument for
  multiple comparisons correction,'' \emph{Neuroimage}, vol.~47, no. Suppl 1,
  p. S125, 2009.

\bibitem{henson2002detecting}
R.~N. Henson, C.~J. Price, M.~D. Rugg, R.~Turner, and K.~J. Friston,
  ``Detecting latency differences in event-related bold responses: application
  to words versus nonwords and initial versus repeated face presentations,''
  \emph{Neuroimage}, vol.~15, no.~1, pp. 83--97, 2002.

\bibitem{aamand2013no}
R.~Aamand, T.~Dalsgaard, Y.-C. {Lynn Ho}, A.~Moller, A.~Roepstorff, and
  T.~Lund, ``A {NO} way to {BOLD}?: Dietary nitrate alters the hemodynamic
  response to visual stimulation,'' \emph{NeuroImage}, vol.~83, 07 2013.

\bibitem{lindquist2009modeling}
M.~A. Lindquist, J.~M. Loh, L.~Y. Atlas, and T.~D. Wager, ``Modeling the
  hemodynamic response function in {fMRI}: efficiency, bias and mis-modeling,''
  \emph{Neuroimage}, vol.~45, no.~1, pp. S187--S198, 2009.

\bibitem{ferrari2012brief}
M.~Ferrari and V.~Quaresima, ``A brief review on the history of human
  functional near-infrared spectroscopy (fnirs) development and fields of
  application,'' \emph{Neuroimage}, vol.~63, no.~2, pp. 921--935, 2012.

\bibitem{ikutani2014brain}
Y.~Ikutani and H.~Uwano, ``Brain activity measurement during program
  comprehension with nirs,'' in \emph{Software Engineering, Artificial
  Intelligence, Networking and Parallel/Distributed Computing}.\hskip 1em plus
  0.5em minus 0.4em\relax IEEE, 2014, pp. 1--6.

\bibitem{Duraes16}
J.~Duraes, H.~Madeira, J.~Castelhano, C.~Duarte, and M.~C. Branco, ``{WAP:
  Understanding the Brain at Software Debugging},'' in \emph{International
  Symposium on Software Reliability Engineering}, 2016, pp. 87--92.

\bibitem{Castelhano2018}
J.~Castelhano, I.~C. Duarte, C.~Ferreira, J.~Duraes, H.~Madeira, and
  M.~Castelo-Branco, ``{The Role of the Insula in Intuitive Expert Bug
  Detection in Computer Code: An {fMRI} Study},'' \emph{Brain Imaging and
  Behavior}, May 2018.

\bibitem{Peitek:2018:ESEM}
N.~Peitek, J.~Siegmund, C.~Parnin, S.~Apel, J.~Hofmeister, and A.~Brechmann,
  ``{Simultaneous Measurement of Program Comprehension with {fMRI} and Eye
  Tracking: A Case Study},'' in \emph{Symposium on Empirical Software
  Engineering and Measurement}, 2018, to appear.

\bibitem{brodmann2007brodmann}
K.~Brodmann, \emph{Brodmann's: Localisation in the cerebral cortex}.\hskip 1em
  plus 0.5em minus 0.4em\relax Springer Science \& Business Media, 2007.

\bibitem{MargulieuxSpatialTheory}
L.~E. Margulieux, ``Spatial encoding strategy theory: The relationship between
  spatial skill and stem achievement,'' in \emph{Proceedings of the 2019 ACM
  Conference on International Computing Education Research}, ser. ICER ’19,
  2019, p. 81–90.

\bibitem{hegarty1999types}
M.~Hegarty and M.~Kozhevnikov, ``Types of visual--spatial representations and
  mathematical problem solving.'' \emph{Journal of educational psychology},
  vol.~91, no.~4, p. 684, 1999.

\bibitem{wai2009spatial}
J.~Wai, D.~Lubinski, and C.~P. Benbow, ``Spatial ability for stem domains:
  Aligning over 50 years of cumulative psychological knowledge solidifies its
  importance.'' \emph{Journal of Educational Psychology}, vol. 101, no.~4, p.
  817, 2009.

\bibitem{sorby2018does}
S.~Sorby, N.~Veurink, and S.~Streiner, ``Does spatial skills instruction
  improve stem outcomes? the answer is ‘yes’,'' \emph{Learning and
  Individual Differences}, vol.~67, pp. 209--222, 2018.

\bibitem{jones2008spatial}
S.~Jones and G.~Burnett, ``Spatial ability and learning to program,''
  \emph{Human Technology: An Interdisciplinary Journal on Humans in ICT
  Environments}, 2008.

\bibitem{parkinson2018investigating}
J.~Parkinson and Q.~Cutts, ``Investigating the relationship between spatial
  skills and computer science,'' in \emph{Proceedings of the 2018 ACM
  Conference on International Computing Education Research}, 2018, pp.
  106--114.

\bibitem{uttal2013malleability}
D.~H. Uttal, N.~G. Meadow, E.~Tipton, L.~L. Hand, A.~R. Alden, C.~Warren, and
  N.~S. Newcombe, ``The malleability of spatial skills: A meta-analysis of
  training studies.'' \emph{Psychological bulletin}, vol. 139, no.~2, p. 352,
  2013.

\bibitem{zacks2008neuroimaging}
J.~M. Zacks, ``Neuroimaging studies of mental rotation: a meta-analysis and
  review,'' \emph{Journal of cognitive neuroscience}, vol.~20, no.~1, pp.
  1--19, 2008.

\bibitem{shepard1971mental}
R.~N. Shepard and J.~Metzler, ``Mental rotation of three-dimensional objects,''
  \emph{Science}, vol. 171, no. 3972, pp. 701--703, 1971.

\bibitem{culham2001neuroimaging}
J.~C. Culham and N.~G. Kanwisher, ``Neuroimaging of cognitive functions in
  human parietal cortex,'' \emph{Current opinion in neurobiology}, vol.~11,
  no.~2, pp. 157--163, 2001.

\bibitem{price2012review}
C.~J. Price, ``A review and synthesis of the first 20 years of pet and fmri
  studies of heard speech, spoken language and reading,'' \emph{Neuroimage},
  vol.~62, no.~2, pp. 816--847, 2012.

\bibitem{vigneau2006meta}
M.~Vigneau, V.~Beaucousin, P.-Y. Herve, H.~Duffau, F.~Crivello, O.~Houde,
  B.~Mazoyer, and N.~Tzourio-Mazoyer, ``Meta-analyzing left hemisphere language
  areas: phonology, semantics, and sentence processing,'' \emph{Neuroimage},
  vol.~30, no.~4, pp. 1414--1432, 2006.

\bibitem{scratch}
\BIBentryALTinterwordspacing
J.~Maloney, M.~Resnick, N.~Rusk, B.~Silverman, and E.~Eastmond, ``The scratch
  programming language and environment,'' \emph{ACM Trans. Comput. Educ.},
  vol.~10, no.~4, Nov. 2010. [Online]. Available:
  \url{https://doi.org/10.1145/1868358.1868363}
\BIBentrySTDinterwordspacing

\bibitem{ParkerSCS1}
M.~C. Parker, M.~Guzdial, and S.~Engleman, ``Replication, validation, and use
  of a language independent cs1 knowledge assessment,'' in \emph{Proceedings of
  the 2016 ACM Conference on International Computing Education Research}, ser.
  ICER ’16, 2016, p. 93–101.

\bibitem{lobato2014impact}
L.~Lobato, J.~M. Bethony, F.~B. Pereira, S.~L. Grahek, D.~Diemert, and M.~F.
  Gazzinelli, ``Impact of gender on the decision to participate in a clinical
  trial: a cross-sectional study,'' \emph{BMC Public Health}, vol.~14, no.~1,
  pp. 1--9, 2014.

\bibitem{sackrowitz1996unlevel}
M.~G. Sackrowitz and A.~P. Parelius, ``An unlevel playing field: Women in the
  introductory computer science courses,'' \emph{ACM SIGCSE Bulletin}, vol.~28,
  no.~1, pp. 37--41, 1996.

\bibitem{huang2019distilling}
Y.~Huang, X.~Liu, R.~Krueger, T.~Santander, X.~Hu, K.~Leach, and W.~Weimer,
  ``Distilling neural representations of data structure manipulation using
  {fMRI} and {fNIRS},'' in \emph{International Conference on Software
  Engineering (ICSE)}, 2019.

\bibitem{nirs-brain-analyzir}
H.~Santosa, X.~Zhai, F.~Fishburn, and T.~Huppert, ``The {NIRS} brain {AnalyzIR}
  toolbox,'' \emph{Algorithms}, vol.~11, no.~5, May 2018.

\bibitem{peters2008applications}
M.~Peters and C.~Battista, ``Applications of mental rotation figures of the
  shepard and metzler type and description of a mental rotation stimulus
  library,'' \emph{Brain and cognition}, vol.~66, no.~3, pp. 260--264, 2008.

\bibitem{GRE}
\BIBentryALTinterwordspacing
ETS.org. (2020) Gre home. ETS. [Online]. Available:
  \url{https://www.ets.org/gre}
\BIBentrySTDinterwordspacing

\bibitem{tew2010developing}
A.~E. Tew and M.~Guzdial, ``Developing a validated assessment of fundamental
  cs1 concepts,'' in \emph{Proceedings of the 41st ACM technical symposium on
  Computer science education}, 2010, pp. 97--101.

\bibitem{barker2013autoregressive}
J.~W. Barker, A.~Aarabi, and T.~J. Huppert, ``Autoregressive model based
  algorithm for correcting motion and serially correlated errors in {fNIRS},''
  \emph{Biomedical optics express}, vol.~4, no.~8, pp. 1366--1379, 2013.

\bibitem{cui2010}
X.~Cui, S.~Bray, and A.~Reiss, ``Functional near infrared spectroscopy ({NIRS})
  signal improvement based on negative correlation between oxygenated and
  deoxygenated hemoglobin dynamics,'' \emph{Neuroimage}, vol.~49, no.~4, pp.
  3039--3046, 2010.

\bibitem{frank-photogram}
X.-S. Hu, N.~Wagley, A.~T. Rioboo, A.~F. DaSilva, and I.~Kovelman,
  ``Photogrammetry-based stereoscopic optode registration method for functional
  near-infrared spectroscopy,'' \emph{Journal of Biomedical Optics}, vol.~25.

\bibitem{cohen1996changes}
M.~S. Cohen, S.~M. Kosslyn, H.~C. Breiter, G.~J. DiGirolamo, W.~L. Thompson,
  A.~Anderson, S.~Bookheimer, B.~R. Rosen, and J.~Belliveau, ``Changes in
  cortical activity during mental rotation a mapping study using functional
  mri,'' \emph{Brain}, vol. 119, no.~1, pp. 89--100, 1996.

\bibitem{cohen2003visual}
L.~Cohen, O.~Martinaud, C.~Lemer, S.~Leh{\'e}ricy, Y.~Samson, M.~Obadia,
  A.~Slachevsky, and S.~Dehaene, ``Visual word recognition in the left and
  right hemispheres: anatomical and functional correlates of peripheral
  alexias,'' \emph{Cerebral cortex}, vol.~13, no.~12, pp. 1313--1333, 2003.

\bibitem{brownsett2010contribution}
S.~L. Brownsett and R.~J. Wise, ``The contribution of the parietal lobes to
  speaking and writing,'' \emph{Cerebral Cortex}, vol.~20, no.~3, pp. 517--523,
  2010.

\bibitem{seghier2013angular}
M.~L. Seghier, ``The angular gyrus: multiple functions and multiple
  subdivisions,'' \emph{The Neuroscientist}, vol.~19, no.~1, pp. 43--61, 2013.

\bibitem{barbey2013dorsolateral}
A.~K. Barbey, M.~Koenigs, and J.~Grafman, ``Dorsolateral prefrontal
  contributions to human working memory,'' \emph{cortex}, vol.~49, no.~5, pp.
  1195--1205, 2013.

\bibitem{kriegeskorte2008representational}
N.~Kriegeskorte, M.~Mur, and P.~A. Bandettini, ``Representational similarity
  analysis-connecting the branches of systems neuroscience,'' \emph{Frontiers
  in systems neuroscience}, vol.~2, p.~4, 2008.

\bibitem{chen2017general}
S.-Y. Chen, Z.~Feng, and X.~Yi, ``A general introduction to adjustment for
  multiple comparisons,'' \emph{Journal of thoracic disease}, vol.~9, no.~6, p.
  1725, 2017.

\bibitem{marks-kovelman-2019}
R.~A. Marks, I.~Kovelman, O.~Kepinska, M.~Oliver, Z.~Xia, S.~L. Haft,
  L.~Zekelman, P.~Duong, Y.~Uchikoshi, R.~Hancock, and F.~Hoeft, ``Spoken
  language proficiency predicts print-speech convergence in beginning
  readers,'' \emph{NeuroImage}, vol. 201, p. 116021, 2019.

\bibitem{ventura2005identifying}
P.~R. Ventura~Jr, ``Identifying predictors of success for an objects-first
  cs1,'' \emph{Computer Science Education}, 2005.

\bibitem{simon1973chess}
H.~A. Simon and W.~G. Chase, ``Perception in chess,'' \emph{Cognitive
  Psychology}, vol.~4, no.~1, pp. 55--81, 1973.

\bibitem{gong2015chess}
\BIBentryALTinterwordspacing
Y.~Gong, K.~Ericsson, and J.~Moxley, ``Recall of briefly presented chess
  positions and its relation to chess skill,'' \emph{PLoS ONE}, vol.~10, no.~3,
  p. e0118756, 2015. [Online]. Available:
  \url{https://doi.org/10.1371/journal.pone.0118756}
\BIBentrySTDinterwordspacing

\bibitem{kovelman2008}
I.~Kovelman, S.~A. Baker, and L.~A. Petitto, ``Bilingual and monolingual brains
  compared: a functional magnetic resonance imaging investigation of syntactic
  processing and a possible ``neural signature'' of bilingualism,''
  \emph{Journal of Cognitive Neuroscience}, vol.~20, no.~1, pp. 153--169, 2008.

\bibitem{lathorp1970}
R.~L. Lathrop, ``How students learn music: The psychology of music and music
  education,'' \emph{Music Educators Journal}, vol.~56, no.~6, pp. 47--145,
  1970.

\bibitem{harris2000selective}
I.~M. Harris, G.~F. Egan, C.~Sonkkila, H.~J. Tochon-Danguy, G.~Paxinos, and
  J.~D. Watson, ``Selective right parietal lobe activation during mental
  rotation: a parametric pet study,'' \emph{Brain}, vol. 123, no.~1, pp.
  65--73, 2000.

\bibitem{bockmon2020cs1}
R.~Bockmon, S.~Cooper, W.~Koperski, J.~Gratch, S.~Sorby, and M.~Dorodchi, ``A
  cs1 spatial skills intervention and the impact on introductory programming
  abilities,'' in \emph{Proceedings of the 51st ACM Technical Symposium on
  Computer Science Education}, 2020, pp. 766--772.

\bibitem{fincher2006predictors}
S.~Fincher, A.~Robins, B.~Baker, I.~Box, Q.~Cutts, M.~de~Raadt, P.~Haden,
  J.~Hamer, M.~Hamilton, R.~Lister \emph{et~al.}, ``Predictors of success in a
  first programming course,'' in \emph{Proceedings of the 8th Australasian
  Computing Education Conference (ACE 2006)}, vol.~52.\hskip 1em plus 0.5em
  minus 0.4em\relax Australian Computer Society Inc., 2006, pp. 189--196.

\bibitem{ParkerSocio}
M.~C. Parker, A.~Solomon, B.~Pritchett, D.~A. Illingworth, L.~E. Marguilieux,
  and M.~Guzdial, ``Socioeconomic status and computer science achievement:
  Spatial ability as a mediating variable in a novel model of understanding,''
  in \emph{Proceedings of the 2018 ACM Conference on International Computing
  Education Research}, ser. ICER '18, 2018, p. 97–105.

\bibitem{cooper15}
\BIBentryALTinterwordspacing
S.~Cooper, K.~Wang, M.~Israni, and S.~Sorby, ``Spatial skills training in
  introductory computing,'' in \emph{International Computing Education
  Research}, 2015, pp. 13--20. [Online]. Available:
  \url{https://doi.org/10.1145/2787622.2787728}
\BIBentrySTDinterwordspacing

\bibitem{xia2017measuring}
X.~Xia, L.~Bao, D.~Lo, Z.~Xing, A.~E. Hassan, and S.~Li, ``Measuring program
  comprehension: A large-scale field study with professionals,'' \emph{IEEE
  Transactions on Software Engineering}, vol.~44, no.~10, pp. 951--976, 2017.

\bibitem{haiduc2010use}
S.~Haiduc, J.~Aponte, L.~Moreno, and A.~Marcus, ``On the use of automated text
  summarization techniques for summarizing source code,'' in \emph{2010 17th
  Working Conference on Reverse Engineering}.\hskip 1em plus 0.5em minus
  0.4em\relax IEEE, 2010, pp. 35--44.

\bibitem{mcburney2015automatic}
P.~W. McBurney and C.~McMillan, ``Automatic source code summarization of
  context for java methods,'' \emph{IEEE Transactions on Software Engineering},
  vol.~42, no.~2, pp. 103--119, 2015.

\bibitem{murphy2012ability}
L.~Murphy, S.~Fitzgerald, R.~Lister, and R.~McCauley, ``Ability to'explain in
  plain english'linked to proficiency in computer-based programming,'' in
  \emph{Proceedings of the ninth annual international conference on
  International computing education research}, 2012, pp. 111--118.

\bibitem{lopez2008relationships}
M.~Lopez, J.~Whalley, P.~Robbins, and R.~Lister, ``Relationships between
  reading, tracing and writing skills in introductory programming,'' in
  \emph{Proceedings of the fourth international workshop on computing education
  research}, 2008, pp. 101--112.

\bibitem{busjahn2011analysis}
T.~Busjahn, C.~Schulte, and A.~Busjahn, ``Analysis of code reading to gain more
  insight in program comprehension,'' in \emph{Proceedings of the 11th Koli
  Calling International Conference on Computing Education Research}, 2011, pp.
  1--9.

\bibitem{prat2020relating}
C.~S. Prat, T.~M. Madhyastha, M.~J. Mottarella, and C.-H. Kuo, ``Relating
  natural language aptitude to individual differences in learning programming
  languages,'' \emph{Scientific reports}, vol.~10, no.~1, pp. 1--10, 2020.

\bibitem{fedorenko2019language}
E.~Fedorenko, A.~Ivanova, R.~Dhamala, and M.~U. Bers, ``The language of
  programming: a cognitive perspective,'' \emph{Trends in cognitive sciences},
  vol.~23, no.~7, pp. 525--528, 2019.

\end{thebibliography}

\end{document}